\documentclass[pra,preprint,twocolumn,10pt,amsmath,amssymb,nofootinbib,bm,bbm,notitlepage,superscriptaddress]{revtex4-2}

\usepackage[english]{babel}
\usepackage[utf8]{inputenc}
\usepackage[T1]{fontenc}

\setcitestyle{sort&compress,numbers}
\usepackage{amsthm, color, mathtools, amsmath, amssymb, setspace,bbm, tensor, subfigure, mathtools, placeins, graphicx, wrapfig,float, verbatim,enumitem}
\usepackage[dvipsnames]{xcolor}
\usepackage[unicode=true, breaklinks=false, pdfborder={0 0 1}, backref=false, colorlinks=true, linkcolor=blue, citecolor=blue]{hyperref}
\usepackage{cancel,bm}
\usepackage{bbm}
\usepackage{orcidlink}
\usepackage{amsbsy}

\let\oldsqrt\sqrt
\def\sqrt{\mathpalette\DHLhksqrt}
\def\DHLhksqrt#1#2{%
\setbox0=\hbox{$#1\oldsqrt{#2\,}$}\dimen0=\ht0
\advance\dimen0-0.2\ht0
\setbox2=\hbox{\vrule height\ht0 depth -\dimen0}%
{\box0\lower0.4pt\box2}}

\newcommand{\tra}{\operatorname{tr}}

\newcommand{\id}{\mathbbm{1}}

\usepackage[normalem]{ulem}

\def\bra#1{\mathinner{\langle{#1}|}}

\def\ket#1{\mathinner{|{#1}\rangle}}

\newcommand*\xbar[1]{%
   \hbox{%
     \vbox{%
       \hrule height 0.5pt 
       \kern0.5ex
       \hbox{%
         \kern-0.2em
         \ensuremath{#1}%
         \kern-0.0em
       }%
     }%
   }%
} 

\def\BraVert{\egroup\,\mid\,\bgroup}

\def\ketbra#1#2{\ket{#1\vphantom{#2}}\!\bra{#2\vphantom{#1}}}

\def\bra#1{\mathinner{\langle{#1}|}}

\def\ket#1{\mathinner{|{#1}\rangle}}

\usepackage{bbm, braket}
\usepackage[mathscr]{euscript}
\allowdisplaybreaks
\newtheorem*{theorem*}{Theorem}

\newtheorem*{corollary*}{Corollary}

\newtheorem*{observation*}{Observation}

\newcommand{\nn}{\nonumber}
\newcommand{\cv}[1]{(#1)}
\newcommand{\cvb}[1]{[#1]}
\newcommand{\cvc}[1]{\{#1\}}
\newcommand{\cvv}[1]{\vert #1\vert}

\newcommand{\cvr}[1]{\left\langle #1\right\rangle}
\newcommand{\iden}{\id}

\newcommand{\oper}[1]{\hat{#1}}

\newcommand{\ds}{\displaystyle}

\DeclareFontFamily{U}{mathx}{}
\DeclareFontShape{U}{mathx}{m}{n}{ <-> mathx10 }{}
\DeclareSymbolFont{mathx}{U}{mathx}{m}{n}
\DeclareFontSubstitution{U}{mathx}{m}{n}
\DeclareMathAccent{\widecheck}{0}{mathx}{"71}

\begin{document}

\title{Noise Decoupling for State Transfer in Continuous Variable Systems}
\author{Fattah Sakuldee\,\orcidlink{0000-0001-8756-7904}}
\email{fattah.sakuldee@ug.edu.pl}
\affiliation{The International Centre for Theory of Quantum Technologies, University of Gda\'nsk, Jana Ba\.zy\'nskiego 1A, 80-309 Gda\'nsk, Poland}
\author{Behnam Tonekaboni\,\orcidlink{0000-0003-4258-3139}}
\email{behnam.tonekaboni@csiro.au}
\affiliation{Quantum Systems, Data61, CSIRO, Clayton, Victoria 3168, Australia}

\begin{abstract}
We consider a toy model of noise channels, given by a random mixture of unitary operations, for state transfer problems with continuous variables. Assuming that the path between the transmitter node and the receiver node can be intervened, we propose a noise decoupling protocol to manipulate the noise channels generated by linear and quadratic polynomials of creation and annihilation operators, to achieve an identity channel, hence the term noise decoupling. For random constant noise, the target state can be recovered while for the general noise profile, the decoupling can be done when the interventions are fast compared to the noise. We show that the state at the transmitter can be written as a convolution of the target state and a filter function characterizing the noise and the manipulation scheme. We also briefly discuss that a similar analysis can be extended to the case of higher-order polynomial generators. 
Finally, we demonstrate the protocols by numerical calculations.
\end{abstract}

\date{\today}
		
\maketitle

\section{Introduction}
Utilizing continuous variables (CV) as tools for quantum computing and manipulations has become one of the promising areas of research in the quantum information community in the past decade \cite{Serafini2017,Braunstein2005}. There are several currently active research aspects in this area, for instance, quantum key distribution \cite{Denys2021,Kanitschar2022,Primaatmaja2022,Srikara2020}, entanglement and resource theory \cite{Srikara2020,Simon2000,Duan2000,Hyllus2006,Mihaescu2020,Kumar2021,Bowen2004}, quantum metrology and states discrimination \cite{Notarnicola2023,Lvovsky2009,Smithey1993,Wang2020,Wu2021}, quantum communication and state transfers \cite{Dequal2021,Rani2023,Notarnicola2023,Lin2020,Bose2023,Anuradha2023,Peuntinger2014}, and noise analysis for CV systems manipulation \cite{Lin2020,Yamano2023,Larsen2020,Walshe2023,Kovalenko2021,Derkach2020,Arenz2017,Lasota2017,Usenko2016,Vitali1999,Bowen2004}.

Despite the different perspectives, there is still room for development and detailed investigation.
For example, it appears that several techniques used in quantum manipulation on finite systems are not translated into a set of tools for controlling CV systems. One of the interesting topics in this picture is the application of noise decoupling and control normally used in solid-state physics \cite{Degen2017,Szankowski2017} to CV systems. The question is \emph{how much one can adapt the tools for noise control from finite systems to CV systems?} This plays an important role in engineering the preparation, measurement, and transfer of information. In this article, we partially contribute by considering a state transfer problem via noisy channels and introducing a decoupling protocol to gain noiseless channels.

{The noise decoupling protocol is a technique of insertion of a series of control operations interlacing a given noisy channel in sequence, to subtract the noise contribution and obtain a clean channel. It is well known in the context of dynamical decoupling for open quantum dynamics where the noise represents the influence from the environment on the system and can be modeled via a unitary evolution of the system-environment} \cite{Viola1998,Viola1999a,Viola1999b}. {In the simplest scenario, where the environmental degrees of freedom are represented by stochastic functions, also known as classical noise representation, the reduced map on the system can be considered as a set of random unitary dynamical maps. Here,} we consider a similar setup with different parameterizations, i.e. using a path distance as a dynamical parameter. For illustration, one can consider communication using photons through fiber or free space, in which the information is encrypted in continuous degrees of freedom of the carrier light \cite{Braunstein2005,Dequal2021}. For instance, the noise in the fiber can be the influence of the medium \cite{Eriksson2019,Bose2023,Villasenor2021}, and the interventions here are simply repeater nodes inserted along the path. In free space communication, even though it is {less practical} to {mid-flight} intervene between transmitter and receiver nodes, e.g. between the ground station and a satellite, a similar picture arises through a protocol of multi-step forward and backward transmissions, in which the first transmitter and the last receiver are treated as first and final nodes of the communication while the intermediate steps {can be considered}  interventions. In these two scenarios, one can expect a protocol of applying control operations at those interventions and obtain a noise decoupling scheme for the CV state transfer. This article aims to find relevant control sets for the simplification models of the aforementioned problem.

The article is organized as follows. Formulations for CV systems and the model of the noise channels are discussed in Sec.~\ref{sec:frame}. In Sec.~\ref{sec:DD}, we recall the basic description of the dynamical decoupling protocol and we discuss the {formulation} of such protocol in our system. 
The decoupling protocol is elaborated in detail for the noisy displacement channels---random noise channels with creation and annihilation generators---both in the filter description in Sec.~\ref{sec:displacement-noise} and in control group averaging picture in Sec.~\ref{sec:DD-picture-for-D}. We further discuss the quadratic generators and the higher-order generators in Sec.~\ref{sec:protocol-squeeze-and-mix}. Numerical illustrations of the protocol are given in Sec.~\ref{sec:numerical}, and finally, the conclusion is given in Sec.~\ref{sec:conclusion}.

\section{Framework}\label{sec:frame}
In this section, we will introduce the setup of our systems and the noise models. We begin with an overview of the noise model and then discuss the mathematical framework and formulation for the noise in our model. We will also recall the Wigner representation {of states} and the {phase space transformations} of the elementary noise operations.

\subsection{Noisy Transfer Channels}\label{sec:noisy-channel}
The basic state transfer (or communication) scheme in quantum information is a protocol to send a {quantum} state $\rho$ on a given Hilbert space $\mathcal{H},$ from one node (transmitter) to another (receiver). The channel through which the transfer has been done is practically modeled as a completely positive and trace preserving (CPTP) transformation $\mathcal{C}$ on a set of bounded operators $\mathcal{B}\cv{\mathcal{H}}$. Then the received state is formally written as $\mathcal{C}\cvb{\rho}.$  In the ideal scenario, such a channel is known and the information inside $\rho$ can be extracted from the received state by standard state tomography \cite{Lvovsky2009}, or parameters characterization when only partial quantities are relevant. 

However, when the communication involves disturbance from the environment or the imperfection of the medium encompassing the channel, the channel becomes partially unknown, resulting in a noisy channel. The simplest model for these channels is ${\mathcal{C}}_{\vec{\lambda}}=\mathcal{E}_{\vec{\lambda}}\circ\mathcal{C},$ where $\mathcal{E}_{\vec{\lambda}}$ is a noise operation {modeled as} a stochastic CPTP map on $\mathcal{B}\cv{\mathcal{H}},$ characterized by a parameters vector $\vec{\lambda}.$ 
We model the noise operation as a random unitary operation equipped with a probability space $(\Omega,\mu)$ in which the map is represented by{
\begin{equation}
    \mathcal{E}_{\lambda} = \int_\Omega d\mu\cv{\omega_{\lambda}}\mathcal{U}_{\omega_{\lambda}},\label{eq:E-int-U}
\end{equation}
where $\mathcal{U}_{\omega_{\lambda}}\cvb{\rho}=\oper{U}_{\omega_{\lambda}}\rho\oper{U}_{\omega_{\lambda}}^\dagger$ on a state $\rho,$ and $\oper{U}_{\omega_{\lambda}}$ is a unitary element for the map $\mathcal{E}_{\lambda}$} (more on that in the next section). {The parameter $\lambda,$ which is one dimensional,} denotes a \emph{path} from the transmitter to the receiver and $\omega_{\lambda}$ is a particular noise configuration for a given path $\lambda.$ 
The main task of noise suppression is to modify the noisy channel ${\mathcal{C}}_{\lambda}$ in order to achieve the noiseless channel $\mathcal{C}.$ 

{For simplicity, let the channel $\mathcal{C}$ be absorbed in the state $\rho,$ the problem then becomes equivalent to manipulating the noise operation $\mathcal{E}_\lambda$ to approach an identity channel $\mathcal{I}.$} In this picture, one simplest example is a communication wire where the parameter $\lambda$ associates the distance from the transmitter node towards the receiver node, and the $\mathcal{E}_\lambda$ is noise at the given position on the transfer path from the origin. 
In the next section, we will consider the structure of this operator in detail.

\subsection{Continuous Variables Systems and Noise Operations}\label{sec:setup}

In this section, we will recall the framework on which we are working in this article. Particularly, we first introduce the continuous variable systems that we are interested in, namely optical systems, and then introduce the noise model.

\subsubsection{Harmonic oscillator as continues variable system}
Let $\mathcal{H}$ be a Hilbert space of square-integrable complex-valued functions on a real line, i.e., \begin{equation}
\mathcal{H}=L^2\cv{\mathbb{R}}=\cvc{\psi:\mathbb{R}\rightarrow\mathbb{C}: \int_{\mathbb{R}}\cvv{\psi\cv{x}}^2dx < \infty}.
\end{equation}
Following conventions in quantum mechanics of continuous variables, we define position {operator} $\oper{x}$ and momentum {operator} $\oper{p}$ on the Hilbert space $\mathcal{H}$ as multiplicative {operator} $\oper{x}\psi\cv{x}{=}x\psi\cv{x}$ and derivative {operator} $\oper{p}\psi\cv{x}{=}-i\partial_x\psi\cv{x}$ respectively. The position and momentum operators follow canonical commutation relation $\cvb{\oper{x},\oper{p}}:=\oper{x}\oper{p}-\oper{p}\oper{x}=i,$ where $i$ is the unit of imaginary numbers. {In terms of position and momentum operators, one can define an annihilation operator $\oper{a}:=\tfrac{1}{\sqrt{2}}(\oper{x}+i\oper{p})$ and a creation operator $\oper{a}^\dagger=\tfrac{1}{\sqrt{2}}(\oper{x}-i\oper{p}),$ and a number basis $\{\phi_n:n=0,1,\ldots\}$ where $\oper{a}\phi_n=\sqrt{n}\phi_{n-1}$ for $ n>0$ and $\oper{a}\phi_0=0$ for a vacuum state $\phi_0.$ Following the canonical commutation relation of position and momentum operators, one has $\cvb{\oper{a},\oper{a}^\dagger}=1.$} We also use a Dirac notation for the states and the dual states, i.e. $\ket{\psi}$ represents the function $\psi$ and $\ket{n}$ represents $\phi_n$ for $n=0,1,\ldots.$

Furthermore, we define a displacement operator $\oper{D}\cv{\alpha}$ and a squeeze operator $\oper{S}\cv{z}$ as follows,
\begin{align}
    \oper{D}\cv{\alpha}&=\exp\cv{\alpha\oper{a}^\dagger - \alpha^*\oper{a}}, \\
    \oper{S}\cv{z}&=\exp\cvb{\cv{z^*\oper{a}^2-z\cv{\oper{a}^\dagger}^2}/2}.
\end{align}
Note that the displacement parameter $\alpha,$ is a complex number where the real part is to \textit{displace} in position space and the imaginary part is to \textit{displace} in momentum space. Similarly, the squeezing parameter $z,$ is a complex number. Here, for the sake of simplicity and without loss of generality, we will mainly discuss real positive squeezing parameters. On the operation level we also write {$\mathcal{D}\cv{\alpha}[\rho]=\oper{D}\cv{\alpha}\rho\oper{D}^\dagger\cv{\alpha}$ and $\mathcal{S}\cv{z}[\rho]=\oper{S}\cv{z}\rho\oper{S}^\dagger\cv{z},$ for a state $\rho.$ Remark here that other displacement and squeeze operators are unitary operators, and when we assign the arguments to be random, these two exemplify the random unitary noise.}

\subsubsection{Noise model}
{We consider noise generated by creation and annihilation operators of degree $2.$ This is a direct generalization of displacement noise and squeezing noise. We model this type of noise by setting} $\omega_\lambda$ {in Eq.~\eqref{eq:E-int-U}} to be a map $\omega_\lambda:\ell\mapsto\cv{\alpha_\ell,z_\ell}\in\mathbb{C}^2$ for $\ell$ lying on the path $\lambda.$  We then consider the unitary elements for the map $\mathcal{E}_\lambda$ of the form
	\begin{align}
		\oper{U}_{\omega_\lambda} &=  \mathcal{T}_{\lambda}\exp\Big[\int_0^{\cvv{\lambda}}\oper{G}\cv{\omega_\lambda\cv{\ell}}d\ell\Big],\nn\\ 
			&=  \mathcal{T}_{\lambda}\exp\Big[\int_0^{\cvv{\lambda}}\oper{G}\cv{\alpha_\ell,z_\ell}d\ell\Big] \label{eq:element-U},
	\end{align}
	where $\cvv{\lambda}$ denotes a path length,
	\begin{align}
		\oper{G}&\cv{\alpha_\ell,z_\ell}\nn\\ &= \Big[\dfrac{1}{2}\cv{z^*_\ell\oper{a}^2-z_\ell\cv{\oper{a}^\dagger}^2}+\cv{\alpha_\ell\oper{a}^\dagger - \alpha^*_\ell\oper{a}}\Big] \label{eq:element-G},
	\end{align}
	for a position $\ell$ on the path $\lambda,$ and $\mathcal{T}_{\lambda}$ is an ordering operator with respect to the path $\lambda$ defined by  
	\begin{equation*}
		\mathcal{T}_{\lambda}\Big[\oper{G}\cv{\omega_\lambda\cv{\ell}}\oper{G}\cv{\omega_\lambda\cv{\ell'}}\Big] = \left\{\begin{array}{lr}
			\oper{G}\cv{\omega_\lambda\cv{\ell}}\oper{G}\cv{\omega_\lambda\cv{\ell'}} ,&\ell\leq\ell'\\
			\oper{G}\cv{\omega_\lambda\cv{\ell'}}\oper{G}\cv{\omega_\lambda\cv{\ell}},&\ell'<\ell.
		\end{array}\right.
	\end{equation*}

For a simple communication wire, one can consider variables $\cv{x,p}$ for the transverse position of a photon carrying a message, and the parameter {$\ell$} indicates the position along the propagating (longitudinal) direction. The action generated by $\oper{G}\cv{\alpha_\ell,z_\ell}$ is a contribution of the noise at the longitudinal position $\ell$ from the propagating origin. Hence $\mathcal{E}_{\omega_\lambda}$ represents the accumulation of noise during the propagation up to the longitudinal position {$\cvv{\lambda}$} on the wire. 

Despite its simplicity, the above model of the noise has been employed to emulate several types of realistic environmental disturbance in optical systems. For instance, the model of noise with random displacement operators has been used to display the detector noise with imperfect efficiency \cite{Lin2020,Yamano2023,Usenko2016}; Gaussian thermal loss \cite{Noh2019}; or an error in Gottesman-Kitaev-Preskill coding \cite{Conrad2023,Lin2023,Fukui2021}. Other phenomena that can be associated with {squeezing noise} include a gate noise \cite{Larsen2020,Walshe2023}; channel noise induced crosstalk \cite{Kovalenko2021}; atmospheric noise in free space communication \cite{Derkach2020}; or the photon loss during a propagation \cite{Lasota2017}. Note that many situations mentioned here may incorporate both displacement and squeezing types. 

\subsection{Wigner Representation and Gaussian Manipulation Framework}
As we explained above, our formulation considers only Gaussian operations, i.e., operations whose generators are at most in the second order of the creation and annihilation operators. We further assume that the state $\rho$ is also given by a mixture of coherent states. It then becomes useful to consider here Wigner functions \cite{Wigner1932,Serafini2017}, where both coherent states (and their mixtures) and Gaussian operations (and their averages) can be illustrated as simple functions and their transformations on phase space. 
In particular, we express {Wigner function of} a density matrix $\rho$ as
	\begin{equation}
		W_{\rho}\cv{x,p} = \frac{1}{2\pi}\int_{-\infty}^\infty dy e^{-ipy}\bra{x+\tfrac{y}{2}}\rho\ket{x-\tfrac{y}{2}} \label{eq:W_def}.
	\end{equation}
We also write the {Wigner} function with a single complex argument as $W_{\rho}\cv{\alpha}=W_{\rho}\cv{x,p}$ where $\alpha=(x+ip)/\sqrt{2}.$ Remark that the Wigner function is normalized, i.e., $\ds\int dx \int dp~\! W_{\rho}\cv{x,p} =1,$ but may be {non-positive.}

In parallel, we employ a Wigner transformation for the operator (a dual Wigner function), namely
	\begin{equation}
		W^G\cv{x,p}=\int_{-\infty}^\infty dy e^{ipy}\bra{x-\tfrac{y}{2}}\oper{G}\ket{x+\tfrac{y}{2}}=W^G\cv{\alpha} \label{eq:W-op-G},
	\end{equation}
for the operator $\oper{G}.$ Here the expectation value of the operator $\oper{G}$ reads
	\begin{align}
		\big\langle\oper{G}\big\rangle_{\rho}&=\tra\cv{\oper{G}\rho}=\int dx\int dp~\! W^G\cv{x,p}W_{\rho}\cv{x,p}\nn\\
			&= \int d\alpha\int d\alpha^*~\! W^G\cv{\alpha}W_{\rho}\cv{\alpha} \label{eq:Bonn-W}.
	\end{align}
For illustration, let us consider a mixture of coherent states $\rho$
    \begin{equation}
	\rho = \sum_rp_r\rho_r\label{eq:rho-def},
    \end{equation}
where $\sum_rp_r=1,$ $p_r>0,$ and {$\rho_r=\ketbra{\beta_r}{\beta_r},$ 
    \begin{align}
	\ket{\beta_r}&=\hat{D}\left(\beta_r\right)\ket{\psi_{\sigma_r}}, \nonumber\\
	\ket{\psi_{\sigma_r}}&=\ds\int_{-\infty}^\infty dx \psi_{\sigma_r}\cv{x}\ket{x}, \nonumber\\
	\psi_{\sigma_r}\left(x\right)&=\bigg(\frac{e^{-x^2/\sigma_r^2}}{2\pi\sigma_r^2}\bigg)^{1/4}, \label{eq:Gaussian-element}
    \end{align}
with a complex number $\beta_r$ indicating the center of the Gaussian state $\rho_r$  on the phase space, equipped with a spreading parameter $\sigma_r^2,$ i.e., a displaced vacuum state of Harmonic oscillator with modified variance. Even though the elements $\rho_r$ are Gaussian, the state Eq.~\eqref{eq:rho-def} itself is not necessarily Gaussian, and its Wigner function reads}
	\begin{align}
		W_{\theta}\cv{x,p} &=\sum_rp_rW_r\cv{x,p},\label{eq:W_rho}\\
		W_r\cv{x,p} &:= \dfrac{1}{\pi}e^{-\frac{\cv{x-\theta_r}^2}{2\sigma_r^2}}e^{-2\sigma_r^2p^2}\label{eq:W_r}.
	\end{align}
{The mixture of Gaussian state Eq.~\eqref{eq:rho-def} is a well-known example of a resourceful class of states called non-Gaussian states \cite{Walschaers2021,Lachman2022}. One notorious example is a Gottesman-Kitaev-Preskill state or a uniform mixture of infinitely many displaced vacuum states with the same inter-distance, which is a promising candidate for fault-tolerant encoding \cite{Gottesman2001}. Remark that our noise operation Eq.~\eqref{eq:E-int-U} with the elements Eq.~\eqref{eq:element-U} can disturb this type of states by deforming the mixture coefficients, resulting in the deterioration of the useful information.}

In this representation, {both displacement and squeeze operators can be associated with transformations of Wigner functions on phase space.} For displacement operator $\oper{D}\cv{\alpha},$ one can define an associate transformation on phase space $T_D\cv{\alpha}$ as
	\begin{equation}
 T_D\cv{\alpha'}\cvb{W_\rho}\cv{\alpha}:=W_{\mathcal{D}\cv{\alpha'}\cvb{\rho}}\cv{\alpha}=W_{\rho}\cv{\alpha+\alpha'} \label{eq:T_D},
	\end{equation}
and similarly the squeeze operator $\oper{S}\cv{z}$ can be associated with $T_S\cv{z}$ via
	\begin{equation}
		T_S\cv{z'}\cvb{W_\rho}\cv{\alpha}:=W_{\mathcal{S}\cv{z'}\cvb{\rho}}\cv{\alpha} \label{eq:T_S}.
	\end{equation}
The exact form of the latter operation is complicated in general, but one of the comprehensive cases is the case when the squeezing parameter $z$ is positive real, {i.e.,} $z=\cvv{z},$ in which the transformation corresponds to scaling of the phase space {axes}. In particular
	\begin{equation}
		T_S\cv{\gamma}\cvb{W_\rho}\cv{x,p}=W_{\mathcal{S}\cv{\gamma}\cvb{\rho}}\cv{x,p} = W_{\rho}\cv{e^{-\gamma} x,e^{\gamma}p} \label{eq:T_S-real},
	\end{equation}
where {$\gamma=\cvv{z}.$} 

The squeeze operators and their transformations are additive in the squeezing parameters, i.e., $T_S\cv{\gamma_1}T_S\cv{\gamma_2}=T_S\cv{\gamma_1+\gamma_2}$ for $\gamma_1$ and $\gamma_2$ real positive. For complex squeezing parameters, one can re-parameterize the squeeze operator into 
	\begin{equation}
		\oper{S}\cv{z}=\oper{S}_{\oper{c}\cv{z}}\cv{\gamma}=\exp\bigg[\dfrac{\gamma}{2}\cv{\oper{c}\cv{z}^2-{\oper{c}^\dagger\cv{z}}^2}\bigg] \label{eq:S_c},
	\end{equation}
where we assume a Bogoliubov-like transformation $\oper{c}\cv{z}=\cv{\sqrt{t}\oper{a}+\sqrt{r}\oper{a}^\dagger}e^{i\theta},$ $t,r>0$ and $t+r=1,$ and we write $z=\gamma\cv{t-r}e^{i\theta}$ with $\gamma>0.$ The corresponding transformation on the phase space similar to Eq.~\eqref{eq:T_S-real} can also be achieved for real coordinates associated with the new operators $\oper{c}\cv{z}$ and $\oper{c}^\dagger\cv{z}.$ However, the additivity property does not hold for the complex parameter. This can be easily seen from the dependency of the parameters in the definition of $\oper{c}\cv{z}$ and $\oper{c}^\dagger\cv{z},$ which can be different for different sections of the path.	

Until now one can observe that in some expressions the representation in real coordinate $\cv{x,p}$ is more transparent than the complex coordinate $\alpha$ and vice versa. Hence, for convenience, we will use real coordinate $\cv{x,p}$ and complex coordinate $\alpha$ interchangeably throughout this paper.

\section{Noise Decoupling Protocols for Displacement Noise}\label{sec:protocol-displace}
Here we recall the general ideas of dynamical decoupling (DD) protocol and briefly discuss the connection to our set-up. We then elaborate on the mechanism for inserting the control operations in our toy model. We discuss the formalism for the displacement noise as manipulation on phase space and then we conclude by revisiting the DD descriptions for that case.

\subsection{Dynamical Decoupling Scheme}\label{sec:DD}
Dynamical decoupling protocols \cite{Viola1998,Viola1999a,Viola1999b} is a well-known control tool for manipulation of signal generated or passed through a system interacting with an environment. The principle is to interlace the dynamical evolution by additional local cyclic unitary operators, leading to an effective Hamiltonian where the system is decoupled from the environment. 

In particular, consider a system $\mathcal{S}$ governed by a {coupled evolution} to an environment $\mathcal{B},$ $\oper{U}_0\cv{t}=e^{-it\oper{H}_0},$ where $\oper{H}_0=\oper{H}_S\otimes\iden+\iden\otimes\oper{H}_B+\oper{H}_{SB}=\sum_\alpha\oper{S}_\alpha\otimes\oper{B}_\alpha$ and {a separable initial state $\rho\cv{0}=\rho_S\cv{0}\otimes\rho_B\cv{0}.$} 
{Note that $\oper{S}_\alpha$ and $\oper{B}_\alpha$ are operators acting locally on the systems and the environment respectively. The dynamical decoupling is a technique of perturbatively inserting a set of cyclic control operators $\oper{U}_1\cv{t},$ equipped with Hamiltonian $\oper{H}_1\cv{t},$ to modify the dynamics in such a way that the effects of the noise are suppressed in the asymptotic limit \cite{Viola1999b}.} To do so, we define a control set $\mathcal{C}_S\subset\mathcal{B}\cv{\mathcal{H}_S}$ --- a set of bounded operators on the system contains $\oper{H}_1\cv{t}$ such that $\oper{U}_1\cv{t}=\ds\mathcal{T}\exp\cvb{-i\int_0^{t}du\oper{H}_1\cv{u}}=\oper{U}_1\cv{t+T_C}$ of a period $T_C,$ where $\mathcal{T}$ is a time order operation. 
It is proven that, for arbitrary interaction $\oper{H}_{SB},$ by putting an arbitrary fast control into the reduced evolution, the average of an observable $\oper{A}\in\mathcal{C}_S$ can be given by \cite{Viola1999a}
	\begin{equation}
		\lim_{N\rightarrow \infty}\tra\cvb{\oper{A}\rho_S\cv{NT_C}} = \tra\cvb{\oper{A}\rho_S\cv{0}}\label{eq:DD-original},
	\end{equation}
where $\rho_S\cv{t}=\tra_B\cvb{\oper{U}_{tot}\cv{t}\rho\cv{0}\oper{U}^\dagger_{tot}\cv{t}},$ and the controlled coupled evolution reads $\oper{U}_{tot}\cv{t}=\ds\mathcal{T}\exp\cvb{-i\int_0^{t}du{H}\cv{u}}$ with ${H}\cv{t}=\sum_\alpha\oper{U}^\dagger_1\cv{t}\oper{S}_\alpha\oper{U}_1\cv{t}\otimes\oper{B}_\alpha.$ If the control set can be arbitrary, { i.e., it covers the whole Hilbert space} $\mathcal{C}_S=\mathcal{B}\cv{\mathcal{H}_S},$ {the expression Eq.~\eqref{eq:DD-original}} can be restated as $\lim_{N\rightarrow \infty}\rho_S\cv{NT_C}=\rho_S\cv{0}.$ {Let $\mathcal{I}_S$ denote an interaction space or a space of all effective local operators on the system $S$ generated by all $\oper{S}_\alpha$ except $\oper{H}_S.$ For a special case, the time integral of the control operators $\oper{U}_1$ reads 
    \begin{equation}    
        \int_0^{T_C}du\oper{U}^\dagger_1\cv{u}\mathcal{I}_S\oper{U}_1\cv{u}=0\label{eq:I_S-zero-ave-int},
    \end{equation}
which is a shorthand notation for  $\ds\int_0^{T_C}du\oper{U}^\dagger_1\cv{u}\oper{A}\oper{U}_1\cv{u}=0$ for all $\oper{A}\in\mathcal{I}_S.$ 
In this case, if the interaction space $\mathcal{I}_S$ is known, one can have}
    \begin{equation}
	\lim_{N\rightarrow \infty}\rho_S\cv{NT_C}=e^{-iNT_C\overline{H}_S}\rho_S\cv{0}e^{iNT_C\overline{H}_S} \label{eq:DD-zero-ave},
    \end{equation}
{where $\overline{H}_S$ is a perturbed local Hamiltonian, i.e., a time average of the local Hamiltonian $\hat{H}_S,$
    \[\overline{H}_S=\dfrac{1}{T_C}\int_0^{T_C}du\oper{U}^\dagger_1\cv{u}\mathcal{H}_S\oper{U}_1\cv{u}.\]
In other words, the influence of the environment on the system is removed asymptotically.}

The procedure above is enabled by two mechanisms. First, the lowest-order Magnus approximation 
    \[\oper{U}_{tot}\cv{NT_C}\approx\exp\cv{-iNT_C\overline{H}_0}= \ds\exp\cvb{-iN\int_0^{T_C}du{H}\cv{u}}\] 
which can be achieved within the given limit. Second, is the symmetry of the average induced by the control set $\mathcal{C}_S.$ This can be seen from writing $\oper{U}_1\cv{t} = \oper{g}_j$ for $t\in[j\Delta t,(j+1)\Delta t)$ and $\Delta t =T_C/\cvv{\mathcal{G}},$ where $\mathcal{G}=\cvc{\oper{g}_j}$ {denotes} a finite group of unitary operators generating $\mathcal{C}_S.$ The lowest order Hamiltonian can then be written as
	\begin{equation}
		\overline{H}_0 = \sum_\alpha \overline{S}_\alpha\otimes\oper{B}_\alpha  =\sum_\alpha \Bigg(\dfrac{1}{\cvv{\mathcal{G}}}\sum_{j}\oper{g}^\dagger_j\oper{S}_\alpha\oper{g}_j\Bigg)\otimes\oper{B}_\alpha\label{eq:H_0}.
	\end{equation}
The group average of an individual term $\overline{S}_\alpha$ commutes with $\mathcal{G}$ and hence $\mathcal{C}_S.$ The dynamical decoupling argument Eq.~\eqref{eq:DD-original} then follows automatically. 
Similarly, {in} the special case, Eq.~\eqref{eq:I_S-zero-ave-int} can be re-expressed in term of group average as
	\begin{equation}
		\dfrac{1}{\cvv{\mathcal{G}}}\sum_{j}\oper{g}^\dagger_j\mathcal{I}_S\oper{g}_j=0 \label{eq:I_S-zero-ave},
	\end{equation}
being a condition for decoupling in Eq.~\eqref{eq:DD-zero-ave}.

{The easiest example of this procedure is the Hahn spin echo experiment where one can decouple the influence of the external field from a spin at the time $\tau$ by inversion at $\tau/2$ \cite{Hahn1950}. For instance, for spin dynamics generated by a Hamiltonian $\oper{H}_\omega=B_\omega\oper{\sigma}_z$ where $B_\omega$ is a static random magnetic field, the dynamical map reads $\oper{U}_\omega(t)=e^{-itB_\omega\oper{\sigma}_z}$ and we observe that the inversion at $\tau/2$ of the evolution $\oper{U}_\omega(\tau)$ leads to
    \begin{equation}
    \oper{\sigma}_x\oper{U}_\omega\!\!\left(\tfrac{\tau}{2}\right)\oper{\sigma}_x\oper{U}_\omega\!\!\left(\tfrac{\tau}{2}\right)=e^{-i(\tfrac{\tau}{2})B_\omega\oper{\sigma}_x\oper{\sigma}_z\oper{\sigma}_x}e^{-i(\tfrac{\tau}{2})B_\omega\oper{\sigma}_z}=\iden \label{eq:Hahn-spin-echo},
    \end{equation}
since $\oper{\sigma}_x\oper{\sigma}_z\oper{\sigma}_x=-\oper{\sigma}_z.$ The second inversion operator $\oper{\sigma}_x$ is used for simplicity. In this example the interaction space $\mathcal{I}_S$ is generated by the operator $\oper{\sigma}_z$ while the control space $\mathcal{C}_S$ is generated by $\oper{\sigma}_x.$ In other words one can consider $\{\iden,\oper{\sigma}_x\}$ a control group (given that the identity operator is a trivial element), and hence the condition  $\oper{\sigma}_x\oper{\sigma}_z\oper{\sigma}_x=-\oper{\sigma}_z$ is simply the group average 
    \[\frac{1}{2}\left(\iden\oper{\sigma}_z\iden+\oper{\sigma}_x\oper{\sigma}_z\oper{\sigma}_x\right)=0,\]
i.e., the group average of the interaction space $\mathcal{I}_S$ vanishes. For a nonstatic magnetic field, one needs to incorporate the perturbation into the formulation as in the standard setup above, and the decoupling can be achieved in the asymptotic limit. Remark here that this also exemplifies the case with classical noise as the environmental degree of freedom is represented as a function, e.g. the magnetic field. } 

For CV systems, there are only a few studies on the application of dynamical decoupling thereon. One of them was done by Vitali and Tombesi in Ref.~\cite{Vitali1999} where they considered a unitary evolution of two oscillators. It is observed that such an evolution can be reversed by a perturbation of a local parity operator on phase space; in this sense, the perturbed Hamiltonian is the original Hamiltonian with a minus sign. Another recent work by Arenz et al. \cite{Arenz2017} contains a detailed description of the generic dynamical {decoupling} protocol for CV systems coupled to another CV system, generalizing the former observation. It is also stated that in general, unlike the finite dimension cases, for arbitrary Hamiltonian, one cannot find a control group/set such that the average concerning it leads to a complete decoupling similar to condition Eq.~\eqref{eq:I_S-zero-ave}. In other words, a control group/set, if there exists, must be designed according to the Hamiltonians, and only some dynamics can be decoupled.

In our case, even though the noise is not described by dynamical evolution, it shares a similar structure to the open dynamics above, where in this case the time parameter $t$ is replaced by the path parameter $\ell.$ The noise in our {case} is considered classical in this perspective since it is represented by a stochastic function depicted in Eq.~\eqref{eq:E-int-U}. In the following, we will elaborate on this picture from the most straightforward scenario where the noise is solely equipped with displacement operators; then we will consider the more complex case of {squeezing noise}; and finally, we will discuss our problem concerning the noise operator of the compound form Eq.~\eqref{eq:element-U}. To do so we need to consider an intervention scenario in our setup.

\subsection{Truncations of Noise Channels and Interventions upon Them}\label{sec:intervention}
\begin{figure}
    \centering
    \includegraphics[width=0.95\columnwidth]{"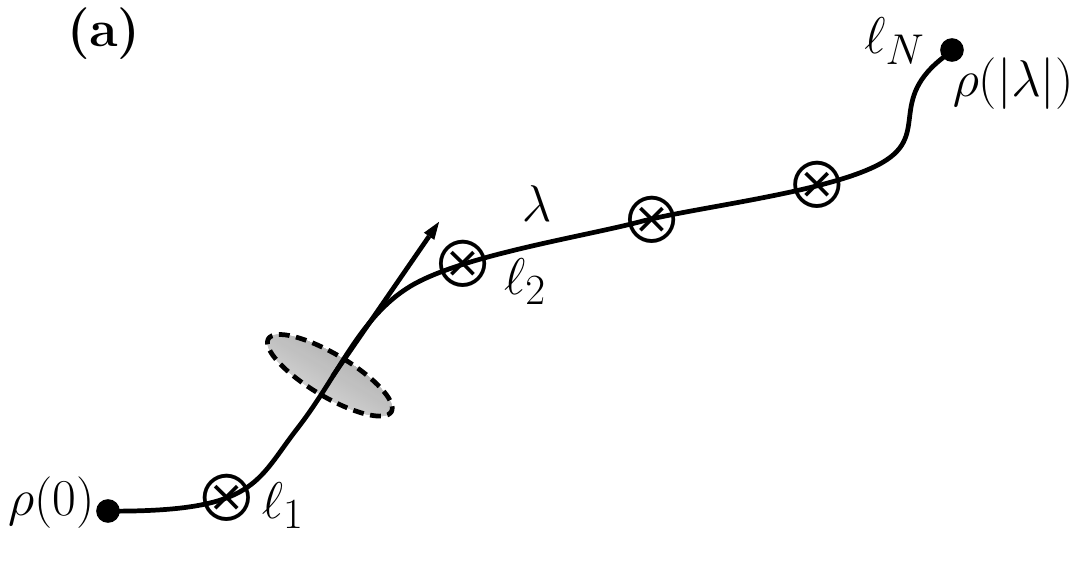"}\\
    \includegraphics[width=0.95\columnwidth]{"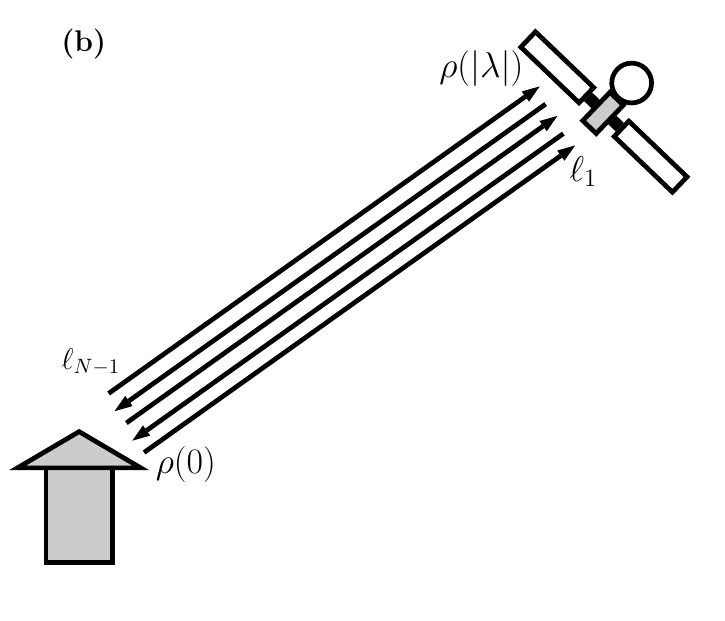"}
    \caption{Schematic of state transfer through a noisy channel with noise decoupling protocols. In {\bf (a)} it is a generic setup of the problem in which the state $\rho(0)$ is transferred from a transmitter node to the receiver node at the other end of the path $\lambda,$ and the receiver will obtain a state $\rho(\vert\lambda\vert).$ {The noise parameter can be described as a complex plane (or more than a single plane) attached to each point on the propagating path,} which can be illustrated as a plane orthogonal to the propagating direction. The circle cross symbols depict the insertions of decoupling operators along the path where their locations are denoted by the path length parameter $\ell.$ When the decoupling operators cannot be placed along the communication path, e.g. the transfer of state from a ground station to a satellite in {\bf (b)}, one can adopt a repeating forward-backward method in which two agents transfer the state back and forth and apply decoupling operators before sending at intermediate time steps. The combined transfers can be considered as a single transfer with noise decoupling intervention.}
    \label{fig:set-up-schematatic}
\end{figure}

Before exploring the manipulation scheme, let us first discuss the physical picture of intervention in our problem. Recall the unitary noise operator $\oper{U}_{\omega_\lambda},$ for a particular noise configuration $\omega_\lambda.$ Let us define a two-point propagator between point $\ell$ to point $\ell'$ on the path $\lambda$ as $\oper{U}_{\omega_\lambda;\ell':\ell}.$ As previously discussed, the path $\lambda$ can be thought of as a transferring path across the communication channel, it is then interesting to consider an additivity concerning the path parameter
	\begin{align}
		\oper{U}_{\omega_\lambda} &= \oper{U}_{\omega_\lambda;\cvv{\lambda}:0} = \oper{U}_{\omega_\lambda;\cvv{\lambda}:\ell}\circ\oper{U}_{\omega_\lambda;\ell:0} \label{eq:U-truncation},
	\end{align}
where $0 \leq \ell\leq \cvv{\lambda}$ and $\oper{U}_{\omega;0:0}=\iden.$ These relations represent the truncation of the channel, for a particular noise configuration $\omega_\lambda$ with respect to the path $\lambda.$ 
One of the simple case for the generator $\oper{U}_{\omega_\lambda;\lambda:0}=\mathcal{T}_\lambda e^{\oper{G}_{\omega_\lambda;\ell':\ell}}$ satisfying the conditions above is 
	\begin{equation}
		\oper{G}_{\omega_\lambda;\ell':\ell}=\int_\ell^{\ell'}\oper{G}_{\omega_\lambda}\cv{s} ds \label{eq:G_ell-ave}
	\end{equation}
for some anti-Hermitian operator $\oper{G}_{\omega_\lambda}\cv{\ell}.$ {Note that we use the symbol $\ell$ as a point on the path $\lambda$ and also the pathlength parameter $0 \leq \ell\leq \cvv{\lambda}.$}

The truncation Eq.~\eqref{eq:U-truncation} is possible by the presence of $\mathcal{T}_\lambda,$ which can be removed when commutativity $\cvb{\oper{G}_{\omega_\lambda}\cv{\ell},\oper{G}_{\omega_\lambda}\cv{\ell'}}=0$ holds for $0\leq\ell,\ell'\leq\cvv{\lambda}.$ These are the special cases of displacement noise ($z_\ell=0$) or real {squeezing noise} ($\alpha_\ell=0$ and $\Im\cv{z}=0$). 
In such two cases, the mentioned divisibility is related to the additivity of displacement operators or squeeze operators with positive real parameters. For more general noise beyond these simple cases, one can remove the symbol $\mathcal{T}_\lambda$ with the help of additional techniques and assumptions, e.g. Trotter product-like approximation and limiting process, which will be discussed later.

By repeating the same mechanism for $n$ steps, we obtain
	\begin{equation}
		\oper{U}_{\omega_\lambda}=\oper{U}_{{\omega_\lambda};\ell_n:\ell_{n-1}}\circ\oper{U}_{{\omega_\lambda};\ell_{n-1}:\ell_{n-2}}\circ\cdots\circ\oper{U}_{{\omega_\lambda};\ell_{1}:0} \label{eq:U-truncation-n},
	\end{equation}
for $\cvv{\lambda}=:\ell_n>\ell_{n-1}>\ldots>\ell_1>0,$ {and} one can define a stochastic process $\vec{\omega}=\cv{\omega_n,\ldots,\omega_1}$ such that 
	\begin{equation}
		\oper{G}_{\omega_k}:=\oper{G}_{\omega;\ell_k:\ell_{k-1}}=\int_{\ell_{k-1}}^{\ell_{k}}\oper{G}_\omega\cv{\ell} d\ell \label{eq:G_omega-k}.
	\end{equation}
Let $\oper{U}_{\omega_k}=\oper{U}_{\omega;\ell_k:\ell_{k-1}}=\mathcal{T}_\lambda e^{\oper{G}_{\omega_k}}$ and $\mathcal{U}_{\omega_k}=\oper{U}_{\omega_k}\cdot\oper{U}^\dagger_{\omega_k}.$ The expression Eq.~\eqref{eq:E-int-U} can then be re-expressed as
	\begin{align}
		\mathcal{E}_{\lambda} &= \int_\Omega d\mu\cv{\omega_{\lambda}}\mathcal{E}_{\omega_\lambda} = \int_{\widetilde{\Omega}} d\tilde{\mu}\cv{\vec{\omega}} \prod_{k=n}^1\mathcal{U}_{\omega_k} \label{eq:E-in-path-sum},
	\end{align}
	where $\tilde{\mu}$ is the distribution of $\vec{\omega}$ on a sample space $\widetilde{\Omega}$ induced by $\mu$ on original space $\Omega.$ 

With the structure given above, we can consider Eq.~\eqref{eq:E-in-path-sum} and discuss a manipulation thereon. 
We aim to perform some engineering over intermediate processes to shape the Eq.~\eqref{eq:E-in-path-sum} to achieve a noiseless one. 
In general, we transcribe such mechanism as additional CPTP maps $\mathcal{A}_k=\sum_{\gamma_k}\hat{A}_{\gamma_k}\cdot\hat{A}^\dagger_{\gamma_k}$ inserted after the action $\mathcal{U}_{\omega_k}$ at each time step $0\leq k\leq n$ and the element $\prod_{k=n}^1\mathcal{U}_{\omega_k}$ will be modified to
	\begin{equation}
		\mathcal{A}_{n}\circ\mathcal{U}_{\omega_n}\circ\mathcal{A}_{n-1}\circ\mathcal{U}_{\omega_{n-1}}\circ\cdots\circ\mathcal{A}_{1}\circ\mathcal{U}_{\omega_1}\circ\mathcal{A}_{0} \label{eq:prod-U-A}.
	\end{equation}
With these at hand, the noise canceling procedure problem is to choose the proper set of operations $\cvc{\mathcal{A}_k}$ (aka control sequence) such that one can approximate the resulting operation Eq.~\eqref{eq:prod-U-A} to an identity operation. 
For an illustration of the setup including the application of intervention operations, see Fig.~\ref{fig:set-up-schematatic}. In the following, we will consider possible control sequences for our particular noise model Eq.~\eqref{eq:element-U}, by beginning with a special case of displacement noise, followed by real {squeezing noise} and an approximation for the general noise from such model.

\subsection{Modification of the Displacement Noise}\label{sec:displacement-noise}
First let us consider the noise when there is only the displacement part in the element Eq.~\eqref{eq:element-U}, i.e. $z=0.$ 
Here the elements for the operations Eq.~\eqref{eq:prod-U-A} of length $n$ reads
	\begin{equation}
		{\mathcal{A}}_{n}\circ\mathcal{D}\cv{\alpha_n}\circ{\mathcal{A}}_{n-1}\circ\mathcal{D}\cv{\alpha_{n-1}}\circ\cdots\circ{\mathcal{A}}_{{1}}\circ\mathcal{D}\cv{\alpha_1}\circ{\mathcal{A}}_{{0}}\label{eq:prod-U-A+D},
	\end{equation}
where $\alpha_k$ corresponds to the process element $\omega_k.$ 
Now define $\mathcal{M}_k$ and $\mathcal{N}_k$ via $\mathcal{A}_k=\mathcal{M}_k\circ\mathcal{N}_k$ for $2\leq k\leq n-1.$ 
We choose the modification operations such that
	\begin{equation}
	 	{\mathcal{N}}_{k}\circ\mathcal{D}\cv{\alpha_{k}}\circ{\mathcal{M}}_{k-1} = \mathcal{D}\cv{s_k\alpha_{k}} \label{eq:A-switch},
	 \end{equation}
for $2\leq k\leq n-1,$ and {write}
	\begin{align}
		{\mathcal{N}}_{1}\circ\mathcal{D}\cv{\alpha_{1}}\circ{\mathcal{A}}_{{0}} &= \mathcal{D}\cv{s_1\alpha_{1}} \label{eq:A-switch-0},\\
		{\mathcal{A}}_{n}\circ\mathcal{D}\cv{\alpha_{n}}\circ{\mathcal{M}}_{n-1} &= \mathcal{D}\cv{s_n\alpha_{n}} \label{eq:A-switch-n}
	\end{align}
where $s_k=\cv{-1}^k$ for $1\leq k\leq n.$ 
It follows that
	\begin{align}
     \mathcal{D}\cv{s_n\alpha_n}\circ\mathcal{D}\cv{s_{n-1}\alpha_{n-1}}\circ\cdots\circ\mathcal{D}\cv{s_1\alpha_1}=\mathcal{D}\bigg(\sum_{k=1}^ns_k\alpha_k\bigg)\label{eq:prod-U-A+D-s}.
    \end{align}
The state after the noise channel will read
	\begin{align}
		\mathbb{E}\cv{\mathcal{E}_{\alpha_\lambda}\cvb{\rho}} &= \mathbb{E}\Bigg[\mathcal{D}\bigg(\sum_{k=1}^ns_k\alpha_k\bigg)\Bigg] \cvb{\rho}\label{eq:prod-U-A+D-s-rho},
	\end{align}
where $\mathbb{E}\cv{X}=\ds\int X~\! d\tilde{\mu}\cv{\alpha_1,\ldots,\alpha_n}$ is a short-hand notation for the average.

Now let us revisit the Wigner representation on phase space. For a state $\rho$ with a Wigner representation $W_{\rho},$ one can have	
	\begin{align}
		T_{\mathcal{D}(\sum_{k=1}^ns_k\alpha_k)}\cvb{W_{\rho}}\cv{\alpha_0} &= W_{\rho}\left(\alpha_0+\sum_{k=1}^ns_k\alpha_k\right) \label{eq:W_AD},
	\end{align}
where the last average is expected to be $W_{\rho}\cv{\alpha_0}$ by our manipulation. 
We define a $2-$dimensional Fourier transform on the phase space 
	\begin{align}
		\mathfrak{F}\cvb{f}\cv{\beta}&=\mathfrak{F}\cvb{f}\cv{\xi,\zeta}\nn\\
			&= \dfrac{1}{2\pi}\int_{-\infty}^\infty dx\int_{-\infty}^\infty dp e^{i\boldsymbol{\beta}^T\boldsymbol{\alpha}}f\cv{\alpha},\nn\\
			&= \dfrac{1}{2\pi}\int_{-\infty}^\infty dx\int_{-\infty}^\infty dp e^{i\cv{x\xi+p\zeta}}f\cv{x,p} \label{eq:F-trans-def},
	\end{align}
$\mathfrak{F}^{-1}$ denotes its inverse transform, and we write $\boldsymbol{\beta}^T\boldsymbol{\alpha}=x\xi+p\zeta$ a scalar product of real vector representations of two complex numbers $\boldsymbol{\alpha}=\cv{x~~p}^T\simeq x+ip$ and $\boldsymbol{\beta}=\cv{\xi~~\zeta}^T\simeq\xi+i\zeta,$ i.e., $\boldsymbol{\beta}^T\boldsymbol{\alpha}=\frac{\beta\alpha^*+\overline{\beta}\alpha}{2}$ in terms of complex numbers operations. Also, 
\[\cvb{f\ast g}\cv{\alpha}=\int d\alpha'\int d\alpha'^*~\!f\cv{\alpha'}g\cv{\alpha-\alpha'}\] 
denotes a ($2-$dimensional) convolution between functions $f$ and $g.$ 
{Employing a shorthand notations for Fourier transform $\widehat{G}\cv{\beta}=\mathfrak{F}\cvb{G}\cv{\beta}=\mathfrak{F}\cvb{G\cv{\alpha}},$ and its inverse transform $\widecheck{G}\cv{\alpha}=\mathfrak{F}^{-1}\cvb{G}\cv{\alpha}=\mathfrak{F}^{-1}\cvb{G\cv{\beta}},$} here the state at the receiver becomes
	\begin{align}
		\mathbb{E}\Bigg[&~\! W_{\rho}\bigg(\alpha_0+\sum_{k=1}^ns_k\alpha_k\bigg)\Bigg] \nn\\
		&= \mathbb{E}\Big[\mathfrak{F}^{-1}\Big( e^{-i\boldsymbol{\beta}^T_0\sum_{k=1}^ns_k\boldsymbol{\alpha}_k} \widehat{W}_{\rho}\cv{\beta_0}\Big)\Big]\nn\\
		&= \dfrac{1}{2\pi}\Big[f_{n}\ast W_{\rho}\Big]\cv{\alpha_0} \label{eq:W_r-conv},
	\end{align}
where  
	\begin{equation}
		f_{n}\cv{\alpha_0} =  \mathbb{E}\Big[\mathfrak{F}^{-1}\Big( e^{-i\boldsymbol{\beta}^T_0\sum_{k=1}^ns_k\boldsymbol{\alpha}_k} \Big)\Big] \label{eq:f_s}. 
	\end{equation}
From Eq.~\eqref{eq:W_r-conv} one can interpret the function $f_{n}$ as a noise filter function where the convolution inside represents a deformation of the target object by the noise and modification we implemented.

At this point let us remark two directions one can handle this issue. First, since the condition for which the perfect noise suppression can be written as
	\begin{equation}
		f_{n}\cv{\alpha}={2\pi}\delta\cv{\alpha}={2\pi}\delta\cv{x}\delta\cv{p} \label{eq:f-delta-cond},
	\end{equation}
the Dirac distribution on the phase space, the direct approach is to modulate the filter function to achieve such a condition. Indeed one can also consider various sequences $s_1,\ldots,s_n$ {apart from what we discussed}, and manipulation thereon, and attempt to modify the convolution function to achieve or at least approximate the delta distribution. Another direction is that, instead of engineering a noise cancellation protocol, one can use the scheme above as a noise spectroscopy, by injecting a known target state $\rho$ and employing availability to adjust $s_1,\ldots,s_n,$ and construct a collection of filter functions $f_{s_1,\ldots,s_n}$. Later one can use such {a collection of filter functions} as a template for the reconstruction of an unknown target state in the original problem. We left the second perspective for further investigation and discussed only the first scenario.

To demonstrate the procedure of noise suppression, we consider an important class of noise distributions e.g. Gaussian noise, where every moment can be characterized by first and second moments \cite{Szankowski2017}. Furthermore, without loss of generality, we assume that the distribution $\mu$ is balanced in such a way that $ \mathbb{E}\cv{\alpha_k} = 0$ for $k=1,\ldots,n,$ hence all odd moments are also $0.$ 
It follows that 
	\begin{align}
		f_{n}&\cv{\alpha_0}= \mathbb{E}\Big[\mathfrak{F}^{-1}\Big( e^{-i\boldsymbol{\beta}^T_0\sum_{k=1}^ns_k\boldsymbol{\alpha}_k} \Big)\Big]\nn\\
		&= \mathfrak{F}^{-1}\Bigg( \exp\bigg\{ -\frac{1}{2}\sum_{kk'} s_ks_{k'}\mathbb{E}\big[\cv{\boldsymbol{\beta}^T_0\boldsymbol{\alpha}_k}\cv{\boldsymbol{\beta}^T_0\boldsymbol{\alpha}_{k'}}\big]\bigg\}\Bigg) \label{eq:f_s-GausNoise-beta},\\
		&= \mathfrak{F}^{-1}\Bigg[ \exp\bigg( -\frac{1}{2}\boldsymbol{\beta}^T_0\boldsymbol{\Sigma}_n\boldsymbol{\beta}_0\bigg)\Bigg] \label{eq:f_s-GausNoise-beta-Matrix},
	\end{align}
where 
	\begin{equation}
		\boldsymbol{\Sigma}_n 
			=\left(\begin{array}{cc}
				A_n & C_n\\
				C_n & B_n
			\end{array} \right)\label{eq:D-Gauss-Sigma-Matrix},
	\end{equation}
	\begin{align}
		A_n &= \dfrac{1}{4}\int_0^{\cvv{\lambda}}d\ell \int_0^{\cvv{\lambda}}d\ell' F_n\cv{\ell}F_n\cv{\ell'}\nn\\ &\hspace{1cm}\times\mathbb{E}\big(\alpha_{\ell}\alpha_{\ell'}+\alpha^*_{\ell}\alpha_{\ell'}+\alpha_{\ell}\alpha^*_{\ell'}+\overline{\alpha_{\ell}\alpha_{\ell'}}\big)\nn\\
			&=\int_0^{\cvv{\lambda}}d\ell \int_0^{\cvv{\lambda}}d\ell' F_n\cv{\ell}F_n\cv{\ell'}\mathbb{E}\cv{x_\ell x_{\ell'}},\label{eq:A-D}\\
		B_n &= -\dfrac{1}{4}\int_0^{\cvv{\lambda}}d\ell \int_0^{\cvv{\lambda}}d\ell' F_n\cv{\ell}F_n\cv{\ell'}\nn\\ &\hspace{1cm}\times\mathbb{E}\big(\alpha_{\ell}\alpha_{\ell'}-\alpha^*_{\ell}\alpha_{\ell'}-\alpha_{\ell}\alpha^*_{\ell'}+\overline{\alpha_{\ell}\alpha_{\ell'}}\big)\nn\\
			&=\int_0^{\cvv{\lambda}}d\ell \int_0^{\cvv{\lambda}}d\ell' F_n\cv{\ell}F_n\cv{\ell'}\mathbb{E}\cv{p_\ell p_{\ell'}},\label{eq:B-D}\\
		C_n &= -\dfrac{i}{4}\int_0^{\cvv{\lambda}}d\ell \int_0^{\cvv{\lambda}}d\ell' F_n\cv{\ell}F_n\cv{\ell'}\mathbb{E}\big(\alpha_{\ell}\alpha_{\ell'}-\overline{\alpha_{\ell}\alpha_{\ell'}}\big)\nn\\
		&= \dfrac{1}{2}\int_0^{\cvv{\lambda}}d\ell \int_0^{\cvv{\lambda}}d\ell' F_n\cv{\ell}F_n\cv{\ell'}\mathbb{E}\big(x_{\ell}p_{\ell'}+p_{\ell}x_{\ell'}\big),\label{eq:C-D}
	\end{align}
and
	\begin{equation}
		F_n\cv{\ell} =
			\left\{\begin{array}{lr}
				-1, & \ell_{k-1}<\ell\leq\ell_{k},~~k\text{~is odd},\\
				1, & \ell_{k-1}<\ell\leq\ell_{k},~~k\text{~is even}.\\
			\end{array}\right. 
			\label{eq:F-Gauss}
	\end{equation}
Suppose that the covariance matrix $\boldsymbol{\Sigma}_n$ can be made positive definite, and the filter function can be reduced to \cite{Serafini2017}
	\begin{equation}
		f_{n}\cv{\alpha_0}= \dfrac{1}{2\pi\sqrt{\det\cv{\boldsymbol{\Sigma}_n}}}\exp\bigg[ -\frac{1}{2}\boldsymbol{\alpha}^T_0\boldsymbol{\Sigma}_n^{-1}\boldsymbol{\alpha}_0\bigg] \label{eq:f-Gauss-normal}.
	\end{equation}
{Note that when the matrix $\boldsymbol{\Sigma}_n$ is not positive definite one can still have filter function Eq.~\eqref{eq:f_s} but will not admit the simple form Eq.~\eqref{eq:f-Gauss-normal}.} With this form at hand, it is transparent that to approximate the perfect decoupling form the noise Eq.~\eqref{eq:f-delta-cond}, one needs to modulate the filter in such a way that $\det\cv{\boldsymbol{\Sigma}_n}$ becomes smaller as $n$ increases, providing a Gaussian smearing for the ideal filter Eq.~\eqref{eq:f-delta-cond}. 

{For illustration, let us consider a noise described by a Compound Poisson Process (CPP) $\alpha(\ell) = \sum_{j=1}^{N(\ell)}\xi_j$ where $\{N(\ell):\ell>0\}$ is a Poisson process of random jump number with a rate $\eta,$ and $\{\xi_j:j\geq 1\}$ obeys identical independent Gaussian distribution (i.i.d. Gaussian) of phase space variable with the mean $0$ and the variance $\sigma_0^2.$ Assuming that the number of interventions $n$ is larger than the number of jumps $N(\cvv{\lambda})$ and assume that the jump only occurs at the intervention point $\ell_k$ on the path $\lambda.$
Let us consider two extreme situations when the number of jumps is equal to the number of interventions, i.e. $N(\cvv{\lambda})=n,$ and the case when there are no jumps at all, i.e. $N(\cvv{\lambda})=0.$ For the first case since $\{\xi_j:j\geq 1\}$ is i.i.d. Gaussian, one can follow the derivation Eqs.~\eqref{eq:f_s}-\eqref{eq:f-Gauss-normal} and obtain 
    \[f_{n}\cv{\alpha_0}=\dfrac{1}{2n(\delta\ell)^2\pi\sigma_0^2}e^{-\cvv{\boldsymbol{\alpha}_0}^2/2n(\delta\ell)^2\sigma_0^2},\]
where we assume that the interval between two adjacent time steps is equal to $\delta\ell=\cvv{\lambda}/n.$ For the second scenario, for an even number of interventions $n$ one can obtain the ideal filter Eq.~\eqref{eq:f-delta-cond} since the term $\sum_{k=1}^ns_k\boldsymbol{\alpha}_k=\boldsymbol{\alpha}_1\sum_{k=1}^ns_k=0$ in Eq.~\eqref{eq:f_s}. For the odd $n,$ one has $\sum_{k=1}^ns_k\boldsymbol{\alpha}_k=\boldsymbol{\alpha}_1,$ and by following the derivation Eqs.~\eqref{eq:f_s}-\eqref{eq:f-Gauss-normal} one obtains
    \begin{equation}
        f_{n}\cv{\alpha_0}=\left\{\begin{array}{lr}
                        2\pi\delta(\alpha_0), & n\text{~is even,~}\\
                        \dfrac{1}{2\pi(\delta\ell)^2\sigma_0^2}e^{-\cvv{\boldsymbol{\alpha}_0}^2/2(\delta\ell)^2\sigma_0^2}, & n\text{~is odd.~} 
                        \end{array}\right. \label{eq:f_s-CPP-const}
    \end{equation}
The second extreme configuration above exemplifies the situation when one can expect a perfect decoupling since the ideal filter Eq.~\eqref{eq:f-delta-cond} can be achieved. 
For the CPP case when the number of jumps is Poisson distributed, one may not expect the ideal filter distribution due to the interference of the structure of the Poisson process. We skip the detailed analysis and consider the related numerical examples in Sec.~\ref{sec:numerical}.} In the following, we revisit the noise decoupling scheme introduced in Sec.~\ref{sec:DD} and show that the principle behind our proposed protocol is simply the same as in the dynamical decoupling scheme.

\subsection{Noise Decoupling Perspective}\label{sec:DD-picture-for-D}
Here let us discuss the noise decoupling for the manipulation scheme above. The introduction of the intervention operations Eq.~\eqref{eq:A-switch} is not arbitrary and the ability to modulate noise to nullity is based on the principle of noise decoupling introduced in Sec.~\ref{sec:DD}. Note that for the displacement noise $\mathcal{D}\cv{\alpha},$ the noise space $\mathcal{I}_S$ (previously called interaction space) is generated by operators $\oper{x}$ and $\oper{p}$ (or equivalently $\oper{a}$ and $\oper{a}^\dagger$). It is clear that, by definitions, all displacement operators are the elements of $\mathcal{I}_S.$ The control space $C_S$ is generated by a group $\mathcal{G}_D=\cvc{\iden,\oper{\Pi}},$ where $\iden$ is an identity operator and $\oper{\Pi}$ is a parity operator defined by $\oper{\Pi}\psi\cv{x}=\psi\cv{-x}$ or $\oper{\Pi}\varphi\cv{p}=\varphi\cv{-p}$ for any wave functions $\psi\cv{x}$ or $\varphi\cv{p}$ \cite{Gerry2010,Chiruvelli2011,Vitali1999}. As an action on the displacement operators, we have 
	\begin{equation}
		\oper{\Pi}\oper{D}\cv{\alpha}\oper{\Pi}=\oper{D}\cv{-\alpha}=\oper{D}^\dagger\cv{\alpha} \label{eq:D_Pi-inversion}
	\end{equation}
for arbitrary $\alpha.$

The group theoretic average of the generators of the displacement noise operators consequently read
	  \begin{equation}
	  	\dfrac{1}{2}\Big(\iden\oper{x}\iden + \oper{\Pi}\oper{x}\oper{\Pi}\Big)=\dfrac{1}{2}\Big(\iden\oper{p}\iden + \oper{\Pi}\oper{p}\oper{\Pi}\Big)=0\label{eq:xp-average}.
	  \end{equation}
In other words,
	\begin{equation}
		\dfrac{1}{\cvv{\mathcal{G}_D}}\sum_{\oper{g}_j\in\mathcal{G}_D}\oper{g}^\dagger_j\mathcal{I}_S\oper{g}_j=0 \label{eq:G_d-ave},
	\end{equation}
which is analogous to the special case in the original dynamical decoupling protocol Eq.~\eqref{eq:I_S-zero-ave}. By this argument, one can have a decoupling scenario analogous to Eq.~\eqref{eq:DD-zero-ave} when one supposes a similar limit to the former expression. {One can notice that the condition Eq.~
\eqref{eq:xp-average} is simply an analog of the group average in the Hanh spin echo experiment Eq.~\eqref{eq:Hahn-spin-echo}. Here the displacement in phase space is similarly inverted by the parity operator as the spin precession direction is inverted in the Hahn spin echo experiment.}

We observe that the manipulation operators in Eqs.~\eqref{eq:A-switch}-\eqref{eq:A-switch-n} are indeed operations associate with elements of $\mathcal{G}_S,$ namely 
	\begin{align}
		\mathcal{N}_k &= \left\{\begin{array}{lr}
			\iden\cdot\iden, &\text{~even~}k,\\
			\oper{\Pi}\cdot\oper{\Pi}, &\text{~odd~}k,
		\end{array}\right.,
		\mathcal{M}_k = \left\{\begin{array}{lr}
			\oper{\Pi}\cdot\oper{\Pi}, &\text{~even~}k,\\
			\iden\cdot\iden, &\text{~odd~}k,
		\end{array}\right.\label{eq:NM-sk-revisit}
	\end{align}
for $k=1,\ldots,n-1,$ $\mathcal{A}_0=\oper{\Pi}\cdot\oper{\Pi}$ and 
	\begin{equation}
		\mathcal{A}_n = \left\{\begin{array}{lr}
			\oper{\Pi}\cdot\oper{\Pi}, &\text{~even~}n,\\
			\iden\cdot\iden, &\text{~odd~}n,
		\end{array}\right.\label{eq:A-n-revisit};
	\end{equation}
in other words, one can simply set
	\begin{equation}
		\mathcal{A}_{k}=\oper{\Pi}\cdot\oper{\Pi} \label{eq:A-sk-revisit}
	\end{equation}
for $k=0,\ldots,n.$ 
In this sense, the state at the receiver Eq.~\eqref{eq:W_r-conv} simply represents a counterpart of evolved state $\rho\cv{NT_C}$ introduced in Sec.~\ref{sec:DD}, where the noise generators $\oper{G}_{\omega}\cv{\ell}$ corresponds to time-dependent Hamiltonians, which is more general than the simple scenario in Sec.~\ref{sec:DD}. 

In the same inscription of $\rho\cv{NT_C},$  one can think of the path parameter $\ell$ as the time parameter $t,$ the average over noise degrees of freedom $\mathbb{E}$ as the partial trace over (classical) bath degrees of freedom, and most importantly the elementary noise operator Eq.~\eqref{eq:U-truncation-n} is analogous to the total time evolution $\oper{U}_{tot}\cv{NT_C}$ where the interventions as in Eq.~\eqref{eq:prod-U-A} exhibit control mechanism. 
Here one can set $T_C=\cvv{\mathcal{G}_D}=2$ and $n=NT_C=2N$ for some $N.$ 
Recall $\oper{G}_{\omega_k}=\alpha_k\oper{a}-\alpha^*_k\oper{a}^\dagger$ for the displacement noise, Eq.~\eqref{eq:prod-U-A+D} can be written as a conjugation of the unitary operator
	{\small
	\begin{equation}
		\exp\bigg(\sum_{m=1}^N\sum_{\oper{g}_k\in\mathcal{G}_D}\oper{g}^\dagger_k\oper{G}_{\omega_{k+2m-1}}\oper{g}_k\bigg)=\exp\bigg(N\sum_{\oper{g}_k\in\mathcal{G}_D}\oper{g}^\dagger_k\overline{G}_{k}\oper{g}_k\bigg) \label{eq:U-D-g-ave},
	\end{equation}}
where $\overline{G}_{k}=\cvr{\alpha_k}\oper{a}-\overline{\cvr{\alpha_k}}\oper{a}^\dagger$ with
	\begin{equation}
		\cvr{\alpha_k} = \frac{1}{N}\sum_{m=1}^N\alpha_{k+2m-1} \label{eq:alpha-steps-ave}.
	\end{equation}
For path parameter independent generators, i.e. $\oper{G}_{\omega}\cv{\ell}=\oper{G}_{\omega}\cv{\ell'}$ for $\ell\neq\ell'$ or consequently $\alpha_k=\alpha_{k'}$ for $k\neq k',$ it is clear that the manipulated noise operator equals to an identity channel for any number $N,$ with the help of Eq.~\eqref{eq:G_d-ave}. Namely, for path-independent displacement noise, the noise can be decoupled and suppressed completely by only one intervention in the middle of the communication path. This resembles the special case Eq.~\eqref{eq:DD-zero-ave} as claimed \footnote{Remark that the limit in Eq.~\eqref{eq:DD-zero-ave} arises from perturbation formalism, while in our case Eq.~\eqref{eq:U-D-g-ave} automatically holds by the {additivity} property of the displacement operators.}. 

For path-dependent generators, although the complete suppression of the noise might not always be the case one can still find situations for which the suppression holds. For instance, assume that the noise is ergodic and $\mathbb{E}\cv{\alpha_k}=\mathbb{E}\cv{\alpha}$ for all $k$ (e.g. stationary), within the limit $N\rightarrow\infty,$ one can claim that the time steps average of the noise parameters Eq.~\eqref{eq:alpha-steps-ave} is identical to configuration average $\mathbb{E}\cv{\alpha}$ \cite{Feller1968,Feller1971}. {As a result,} one can say that
	\begin{equation}
		\lim_{n\rightarrow\infty}\prod_{k=1}^n\mathcal{A}_k\circ\mathcal{D}\cv{\alpha_k}=\mathcal{I} \label{eq:D-to-I-ergodic}
	\end{equation}
for $\mathcal{A}_k$ is given in Eq.~\eqref{eq:A-sk-revisit} and when the noise is ergodic and stationary. 

Finally, we note that the similar analysis above can also be extended to a more generic case with general interventions Eq.~\eqref{eq:prod-U-A}. In particular, by setting $T_C=\cvv{G}$ for some control algebra such that the relation Eq.~\eqref{eq:G_d-ave} holds and setting $n=NT_C=N\cvv{G},$ the manipulated noise operation in such expression can be described as a conjugation of
	\begin{equation}
		\mathcal{T}_\lambda\exp\Bigg(N\sum_{\oper{g}_k\in\mathcal{G}_D}\oper{g}^\dagger_k\overline{G}_{\omega_k}\oper{g}_k\Bigg) \label{eq:U-g-ave},
	\end{equation}
where $\overline{G}_{\omega_k} = \frac{1}{N}\sum_{m=1}^N\oper{G}_{\omega_k}.$ 
Here, by assuming that the noise is ergodic and $\int_{\widetilde{\Omega}} d\tilde{\mu}\oper{G}_{\omega_k}=\int_{\widetilde{\Omega}} d\tilde{\mu}\oper{G}_{\omega}$ for all $k$ (e.g. stationary), it follows that $\lim_{N\rightarrow\infty}\overline{G}_{\omega_k}=\int_{\widetilde{\Omega}} d\tilde{\mu}\oper{G}_{\omega}$, and by the vanishing of average noise operator Eq.~\eqref{eq:G_d-ave}, one could expect that
	\begin{equation}
		\lim_{n\rightarrow\infty}\Bigg(\prod_{k=1}^n\mathcal{A}_k\circ\mathcal{U}_{\omega_k}\Bigg)\circ\mathcal{A}_0=\mathcal{I} \label{eq:U-to-I-ergodic}.
	\end{equation} 
The next section will discuss this scenario, as well as its related spectroscopy filters, for the {squeezing noise} and the combined noise introduced in Sec.~\ref{sec:setup}.

\section{Noise Decoupling Protocols for Squeezing Noise and Combine Noise}\label{sec:protocol-squeeze-and-mix}
We have demonstrated the implementation of the DD protocol for the case of displacement noise. We will show that a similar picture can be obtained for the {squeezing noise} with real parameters. We further illustrate that, under the condition that the control operations can be inserted arbitrarily fast and the noise is stationary, one can also obtain a complete decoupling {in the asymptotic limit} for the generic {squeezing noise} and the noise generated by any quadratic polynomial in creation and annihilation operators. Lastly, we give an overview of a possible extension of the protocol to the case for the noise operators generated by a polynomial in creation and annihilation operators of a degree higher than two.

\subsection{Manipulation on Squeezing Noise with Real Positive Squeezing Parameters}\label{sec:squeezing-noise-real}
Now let us consider the cases when there is only {squeezing noise}, where we propose a control set inspired by the noise decoupling perspective and then discuss the manipulation in Wigner representation including filtering mechanism and noise spectroscopy. 
We begin with the average of noise space $\mathcal{I}_S,$ which is, in this case, generated by $\oper{a}^2$ and $\cv{\oper{a}^\dagger}^2.$ Now we consider
	\begin{equation}
		\mathcal{G}_S=\cvc{\iden,\oper{R}_{\frac{\pi}{2}}=e^{-i\pi\oper{a}\oper{a}^\dagger/2}}\label{eq:G_S-squeeze}
	\end{equation}
as a control set \footnote{In principle, to make a clear comparison to the noise decoupling scheme, one should consider a cyclic group $\cvc{\iden,\oper{R}_{\frac{\pi}{2}},\oper{R}_{\frac{3\pi}{2}},\oper{R}_{\pi}}$. Although the set $\mathcal{G}_S$ is not a cyclic group, it suffices to consider this set for the control of squeeze operators since the squeeze operators have two-fold symmetry, making the action of $\oper{R}^2_{\frac{\pi}{2}}$ similar to that of $\iden,$ and $\oper{R}^3_{\frac{\pi}{2}}$ to $\oper{R}_{\frac{\pi}{2}},$ respectively.}. The average of the noise space for the set $\mathcal{G}_S$ is zero, i.e.
	\begin{equation}
		\dfrac{1}{\cvv{\mathcal{G}_S}}\sum_{\oper{g}_j\in\mathcal{G}_D}\oper{g}^\dagger_j\mathcal{I}_S\oper{g}_j=0 \label{eq:G_S-ave},
	\end{equation}
since for a real number $\omega,$
	$e^{i\omega\oper{a}\oper{a}^\dagger}\oper{a}e^{-i\omega\oper{a}\oper{a}^\dagger}=e^{i\omega}\oper{a};$
or
	\begin{equation}
	  	\dfrac{1}{2}\Big(\iden\oper{a}^2\iden + \oper{R}^\dagger_{\frac{\pi}{2}}\oper{a}^2\oper{R}_{\frac{\pi}{2}}\Big)=\dfrac{1}{2}\Big(\iden\cv{\oper{a}^\dagger}^2\iden + \oper{R}^\dagger_{\frac{\pi}{2}}\cv{\oper{a}^\dagger}^2\oper{R}_{\frac{\pi}{2}}\Big)=0\label{eq:squeeze-gen-average}.
	  \end{equation}
In other words, the control set $\mathcal{G}_S$ can be used to modify {squeezing noise}, and a similar decoupling scheme within the same spirit can be formulated as in the displacement noise.

From the operation picture, similar to the action of $\oper{\Pi},$ which reverses displacement operators, the rotation $\oper{R}_{\frac{\pi}{2}}$ does reverse squeeze operators, i.e.,
	\begin{equation}
		\oper{R}^\dagger_{\frac{\pi}{2}}\oper{S}\cv{z}\oper{R}_{\frac{\pi}{2}}=\oper{S}\cv{-z}=\oper{S}^\dagger\cv{z}\label{eq:S_R-inversion},
	\end{equation}
regardless of the parameter $z.$ We will use this identity to construct a filter similar to the case of displacement noise. To do so, for the {squeezing noise} with manipulation
	\begin{equation}
		{\mathcal{A}}_{n}\circ\mathcal{S}\cv{z_n}\circ{\mathcal{A}}_{n-1}\circ\mathcal{S}\cv{z_{n-1}}\circ\cdots\circ{\mathcal{A}}_{{1}}\circ\mathcal{S}\cv{z_1}\circ{\mathcal{A}}_{{0}}\label{eq:prod-U-A+S},
	\end{equation}
one can set	
	\begin{equation}
		\mathcal{A}_{k} =\left\{\begin{array}{lr}
			\oper{R}_{\frac{\pi}{2}}\cdot\oper{R}^\dagger_{\frac{\pi}{2}}, &\text{~even~}n,\\
			\oper{R}^\dagger_{\frac{\pi}{2}}\cdot\oper{R}_{\frac{\pi}{2}}, &\text{~odd~}n,
		\end{array}\right.,\label{eq:A-k-S}
	\end{equation}
for $k=1,\ldots,n,$ and $\mathcal{A}_0=\oper{R}_{\frac{\pi}{2}}\cdot\oper{R}^\dagger_{\frac{\pi}{2}}.$ Specifically, in parallel with the displacement noise, we define $\mathcal{M}_k$ and $\mathcal{N}_k$ via $\mathcal{A}_k=\mathcal{M}_k\circ\mathcal{N}_k$ for $2\leq k\leq n-1,$ and choose 
	\begin{align}
		\mathcal{N}_k &= \left\{\begin{array}{lr}
			\iden\cdot\iden, &\text{~even~}n,\\
			\oper{R}^\dagger_{\frac{\pi}{2}}\cdot\oper{R}_{\frac{\pi}{2}}, &\text{~odd~}n,
		\end{array}\right.,
		\mathcal{M}_k = \left\{\begin{array}{lr}
			\oper{R}_{\frac{\pi}{2}}\cdot\oper{R}^\dagger_{\frac{\pi}{2}}, &\text{~even~}n,\\
			\iden\cdot\iden, &\text{~odd~}n,
		\end{array}\right.\label{eq:NM-S-sk-revisit}
	\end{align}
for $k=1,\ldots,n-1,$ {and,} $\mathcal{A}_0$ and $\mathcal{A}_n$ are defined as above.
Finally, we arrived at
	\begin{equation}
		\mathcal{S}\cv{s_nz_n}\circ\mathcal{S}\cv{s_{n-1}z_{n-1}}\circ\cdots\circ\mathcal{S}\cv{s_1z_1}\label{eq:prod-U-A+S-s},
	\end{equation}
where $s_k=\cv{-1}^k$ for $k=1,\ldots,n.$

However, unlike the displacement noise, without further assumption, the Eq.~\eqref{eq:prod-U-A+S-s} cannot be combined trivially into a squeeze operation with a parameter comprising the parts. Hence from now on, we focus on real positive parameters $z_k=\gamma_k>0,$ and later will consider an approximation for general squeezing parameters within an appropriate limit. We write
	\begin{equation}
		\mathcal{S}\cv{s_n\gamma_n}\circ\mathcal{S}\cv{s_{n-1}\gamma_{n-1}}\circ\cdots\circ\mathcal{S}\cv{s_1\gamma_1}=\mathcal{S}\bigg(\sum_{k=1}^ns_k\gamma_k\bigg)\label{eq:prod-U-A+S-gamma-s}.
	\end{equation}
Clearly, for the path-independent noise, only one intervention or $n=2$ suffices to decouple the noise from the state at the receiver. For the path-dependent noise, without limiting procedure, it leads to a convolution between the target state and a filter function constituted by noise operations and our modification operations.

Recall that 
	\begin{align}
		T_{\mathcal{S}(\sum_{k=1}^n\gamma_k)}\cvb{W_{\rho}}\cv{\alpha_0} &= W_{\rho}\left(e^{-\Gamma_n}x_0,e^{\Gamma_n}p_0\right) \label{eq:W_AS},
	\end{align}	
where $\Gamma_n=\sum_{k=1}^ns_k\gamma_k.$ {For the filter picture} in this case one may write
	\begin{align}
		\mathbb{E}\Bigg[&~\! W_{\rho}\left(e^{-\Gamma_n}x_0,e^{\Gamma_n}p_0\right)\Bigg] &= \dfrac{1}{2\pi}\Big[f_{n}\ast W_{\rho}\Big]\cv{x_0,p_0} \label{eq:W_r-conv-S},
	\end{align}
where for $x\neq 0$ and $p\neq 0,$
	\begin{equation}
		f_{n}\cv{x,p} = \ds\int_{-\infty}^\infty\int_{-\infty}^\infty 
		e^{-i\cv{r\ln\cvv{x}+t\ln\cvv{p}}}\mathbb{E}\big[e^{i\Gamma_n\cv{r-t}}\big] drdt\label{eq:f_s-S},
	\end{equation}
otherwise, we write
    \begin{equation}
	f_{n}\cv{x,p} = \left\{\begin{array}{lr}
			\ds\int_{-\infty}^\infty e^{-it\ln\cvv{p}}\mathbb{E}\big[e^{i\Gamma_nr}\big] dt, & x= 0 \text{~and~}p\neq 0,\\
			\ds\int_{-\infty}^\infty e^{-ir\ln\cvv{x}}\mathbb{E}\big[e^{-i\Gamma_nt}\big] dr, & x\neq 0 \text{~and~}p= 0,\\	
			1, & x=p=0.	
		\end{array}\right. \label{eq:f_s-S-other}
	\end{equation}
The form of the filter function here is obtained simply by changing variables from $\cv{x,p}$ to $\cv{u,v}=\cv{\ln\cvv{x},\ln\cvv{p}}$ for $x\neq 0$ and $p\neq 0,$ following by performing Fourier transform concerning new variables and using the same analysis as in the case of displacement noise. We skip the detailed analysis of the filter functions and revisit the cases of non-additive noise.

\subsection{Manipulation on Squeezing Noise with Complex Squeezing Parameters}\label{sec:squeezing-noise-complex}
For {squeezing noise} with complex parameters, as previously discussed, one can no longer employ the additivity of the actions on phase space in the case of positive real parameters. To overcome this issue one needs to consider large $n$ limit, or equivalently when the quantity $\delta\ell=\max_{1\leq k\leq n}\cv{\ell_k-\ell_{k-1}}$ approaches zero. With this one can sloppily suppose a Trotter-like limit for Eq.~\eqref{eq:prod-U-A+S-s} as {
	\begin{equation}
		\prod_{k=1}^n\mathcal{S}\cv{s_kz_k}\approx\mathcal{S}\bigg(\sum_{k=1}^ns_kz_k\bigg)\label{eq:prod-U-A+S-s-trotter},
	\end{equation}
for large $n.$}
Following this one can derive a filter for the noise similar to the case of positive squeezing parameters.

In general, the relation Eq.~\eqref{eq:prod-U-A+S-s-trotter} may not hold since the operators involved in the actions are unbounded. However, the range of the parameters is controlled by the distribution $\tilde{\mu},$ which is usually supposed to concentrate in some relevant region, and provides a reasonable approximation for the relation Eq.~\eqref{eq:prod-U-A+S-s-trotter}. 
Roughly speaking, by setting $R=\max_{1\leq k\leq n}\cvv{z_k},$ the {difference between the left and right hand sides in} Eq.~\eqref{eq:prod-U-A+S-s-trotter} is controlled by $\mathcal{O}\bigg\{\exp\big[R^2\cv{\delta\ell}^2\big]\bigg\}.$ The contribution from large $R^2$ is suppressed by the small mass $\tilde{\mu}$ around its region while the the contribution $\cv{\delta\ell}^2$ vanishes within the limit. 
In other words, for {squeezing noise} with complex parameters, it is possible to choose a set of parameters to achieve the filter picture for the noise decoupling scheme via Eq.~\eqref{eq:prod-U-A+S-s-trotter}, at least in the sense of approximation.

\subsection{Displacement and Squeezing Combine Noise and Observations on Non-Gaussian Noise}\label{sec:protocol-mix-noise}
Now let us {briefly} discuss the condition for the generic noise in our model Eq.~\eqref{eq:element-U}. First, recall the form of generator Eq.~\eqref{eq:element-G}
	\begin{align}
		\oper{G}&\cv{\alpha_\ell,z_\ell}\nn\\ &= \Big[\dfrac{1}{2}\cv{z^*_\ell\oper{a}^2-z_\ell\cv{\oper{a}^\dagger}^2}+\cv{\alpha_\ell\oper{a}^\dagger - \alpha^*_\ell\oper{a}}\Big] \label{eq:element-G-recall}.
	\end{align}
It can be seen that the noise space $\mathcal{I}_S$ is generated by $\cvc{\oper{a},\oper{a}^\dagger,\cv{\oper{a}}^2,\cv{\oper{a}^\dagger}^2}.$ 
We define
	\begin{equation}
		\mathcal{G} = \cvc{\iden,\oper{R}_{\frac{\pi}{2}},\oper{R}_{\pi},\oper{R}_{\frac{3\pi}{2}}} \label{eq:Gaussian-G},
	\end{equation}
leading to
	\begin{equation}
		\dfrac{1}{\cvv{\mathcal{G}}}\sum_{\oper{g}_j\in\mathcal{G}}\oper{g}^\dagger_j\mathcal{I}_S\oper{g}_j=0 \label{eq:Gaussian-G-I_S-ave},
	\end{equation}
and hence $\mathcal{G}$ can be termed as a control group for the generic noise in our model. Note that control operators in the cases of displacement noise and {squeezing noise} are also generated by $\mathcal{G}.$ By employing the control operators from the group $\mathcal{G}$ and using the limiting procedure similar to Eq.~\eqref{eq:prod-U-A+S-s-trotter}, one will be able to formulate a similar analysis as in the displacement noise, e.g. {noise decoupling} picture Eq.~\eqref{eq:U-to-I-ergodic} or the filter for noise spectroscopy Eq.~\eqref{eq:W_r-conv}.

Lastly, let us consider a possible extension of the noise beyond the second order in creation and annihilation operators---non-Gaussian noise. 
For Gaussian noise (our model), one can see that the control space is simply generated by a cyclic group of rotation with fundamental angle $\dfrac{\pi}{2}.$ Now we show that one can construct a control group for non-Gaussian noise also from this principle. In particular, let us consider a noise operator Eq.~\eqref{eq:element-U} with the replacement of the elementary generators by
	\begin{equation}
		\oper{G}\cv{b_m,\ldots,b_1}= \sum_{k=1}^m\big[b_k\cv{\ell}{\oper{a}}^k+{b}^*_k\cv{\ell}\cv{\oper{a}^\dagger}^k\big] \label{eq:element-G-non-Gauss},
	\end{equation}
a degree $m$ polynomial in the creation and annihilation operators with (random) complex coefficients $b_k\cv{\ell}$ and ${b}^*_k\cv{\ell}$ for $k=1,\ldots,m.$ The noise space in this case is generated by $\cvc{{\oper{a}}^k,\cv{\oper{a}^\dagger}^k}_{k=1}^m.$ Now we set $\oper{g}_1=\oper{R}_{\frac{\pi}{m}}=e^{-i\pi\oper{a}\oper{a}^\dagger/m}$ and the group  
        \begin{equation}
            \mathcal{G}=\cvc{\oper{g}_1^j=\oper{R}_{\frac{j\pi}{m}}}_{j=0}^{2m-1} \label{eq:G-group-non-Gaussian},
        \end{equation}
generated by $g_1.$ Since $\oper{g}^\dagger_1\oper{a}\oper{g}_1=e^{i\pi/m}\oper{a},$ it follows that
	\begin{equation*}
		\sum_{j=0}^{2m-1}\cv{\oper{g}^\dagger_1}^j\cv{\oper{a}}^p\oper{g}_1^j  
		= \Bigg(\dfrac{1-e^{2ip\pi}}{1-e^{ip\pi/m}}\Bigg)\oper{a}^p=0
	\end{equation*}
for $p\leq m.$ Hence, we will have
	\begin{equation}
		\dfrac{1}{\cvv{\mathcal{G}}}\sum_{\oper{g}_j\in\mathcal{G}}\oper{g}^\dagger_j\mathcal{I}_S\oper{g}_j=0 \label{eq:non-Gaussian-G-I_S-ave},
	\end{equation}
or the group $\mathcal{G}$ here can be considered as a control group for the noise with the generators Eq.~\eqref{eq:element-G-non-Gauss} as claimed.

\section{Numerical Examples}\label{sec:numerical}

In this section, we test our protocol numerically. We consider a channel with total length $|\lambda|$  and numerically calculate the dynamics of $\rho(\ell)$, for $\ell \in [0,|\lambda|]$, under the effect of noise and the external control intervention. Initializing our system in a pure state, results in a mixed state $\rho(\ell)$ due to the noise. To quantify the effect of the noise, we calculate the fidelity of $\rho(\ell)$ compared to $\rho(0)$ via
\begin{align}
    F(\ell):&= F(\rho(\ell), \rho(0)) \notag \\
            &=  \left(\mathrm{Tr} \sqrt{\sqrt{\rho(\ell)} \rho(0) \sqrt{\rho(\ell)}} \right)^2.
\end{align}
In the ideal case, i.e., in the absence of noise, $\rho(\ell)=\rho(0)$ and the fidelity is $1$ for all $\ell$. Because noise reduces fidelity, the aim is to demonstrate that our control protocol results in higher fidelity than the no-control scenario. 

\begin{figure}[t]
\includegraphics[width=0.95\columnwidth]{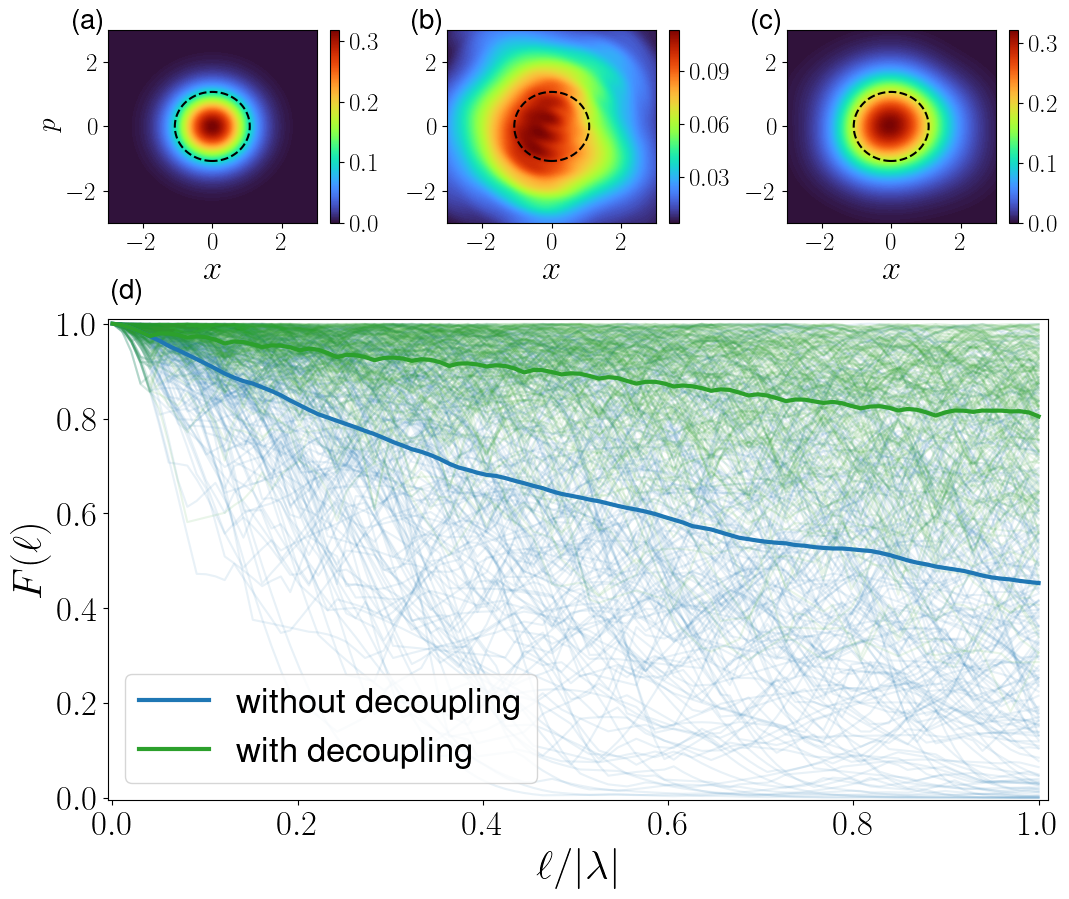}
\caption{\label{fig:DisNoise} {Numerical} demonstration of noise suppression of random displacement. The top row shows the Wigner function of (a) the vacuum state, the initial state at $\ell=0$, (b) the {average of $200$ trajectories of the} final state in the lack of intervention at $\ell=|\lambda|$, and (c) the {average of $200$ trajectories of the} final state with the intervention at $\ell=|\lambda|$. {Note that the asymmetry in (b) is due to the finite number of trajectories.} The dashed circle is a contour for the vacuum state where the  Wigner function is at $10$ percent of its maximum. We re-sketched the same circle on (b) and (c) for ease of comparison. (d) shows the Fidelity between the state at $\ell/|\lambda|$ and initial state. Blue color represents fidelity without intervention and green represents intervention. Thin lines are for individual trajectories of noise and the blue(green) thick line shows the average Fidelity without(with) intervention.}
\end{figure}

{For the sake of simplicity,} we choose that the initial state is a vacuum state with density operator $\rho(0) = \rho_{\mathrm{vac}} =  \ket{0}\bra{0}$; this state represents the ground state of a harmonic oscillator. The Wigner function of $\rho_\mathrm{vac}$ is a symmetric Gaussian function with standard deviation $\sigma_0$ as 
\begin{equation}
	W_{\rho_\mathrm{vac}}\cv{x,p} = \dfrac{1}{\pi}e^{-\frac{x^2+p^2}{2\sigma_0^2}}.
\end{equation}

In the following subsections, we will look at displacement noise, {squeezing noise}, and, the combination of both. We quantitatively examine our control protocol in each instance.

\subsection{Displacement Noise}\label{sec:numer-displace}
In this subsection, we consider that the noise is only a random displacement noise $\oper{D}(\zeta)$, such that by knowing the state $\rho(\ell)$, we calculate the state at $\ell + \mathrm{d}\ell$ via 
\begin{equation}\label{eq:dis_state}
    \rho(\ell + \mathrm{d}\ell) = \oper{D}(\zeta) \rho(\ell) \oper{D}(\zeta)^\dagger,
\end{equation}
where $\zeta$ is a random complex number; $\zeta = \zeta_r + i \zeta_j$. We model the noise by taking $\zeta_r$ and $\zeta_j$ to be random real values from a Compound Poisson Process (CPP). The use of CPP entails that we assume $\zeta$ jumps between different values and is constant between jumps. In terms of Wigner function dynamics, this means that the Wigner function moves in the direction of $\zeta$ for some time and a jump in $\zeta$ is associated with changing direction. Moreover, CPP is defined by a rate $\eta$ that determines how quickly the values of $\zeta_r$ and $\zeta_i$ change, and a probability distribution function (pdf) from which new $\zeta$'s are chosen. In our numerical simulation, we employ a Gaussian pdf with a standard deviation of $\sigma \le \sigma_0$. Furthermore, we choose a slowly varying noise with $\eta \le 0.3$.

The effect of such displacement noise on a vacuum state is shown in Figure~\ref{fig:DisNoise}, where its sub-figure (a) shows the Wigner function of the vacuum state and the dashed circle is a contour where the Wigner function is at $10$ percent of its maximum. Sub-figure (b) shows the Wigner function of average density matrix $\overline\rho(|\lambda|)$ where the average density matrix is defined as
{\begin{equation}
    \overline\rho(\ell) = \sum_{\zeta=1}^{M} \rho_{\zeta}(\ell).
\end{equation}}
Here, in our numerical calculation, we choose $\eta=0.2$ and $\sigma=0.05$. We averaged over {$M=200$} noise realizations. The thin blue lines in Figure~\ref{fig:DisNoise}(d) show the fidelity of $\rho(\ell)$, compared to $\rho(0)$ for {$200$} noise trajectories and the thick blue line is the average Fidelity. As a result, the numerical calculation shows a rapid drop in fidelity due to random displacement noise. 

To suppress the noise, we use our protocol and apply multiple interventions, such that in the presence of an intervention, Eq.~\eqref{eq:dis_state} becomes
\begin{equation}
    \rho(\ell + \mathrm{d}\ell) = \oper{D}(\zeta) \oper{A} \rho(\ell) \oper{A}^\dagger \oper{D}(\zeta)^\dagger
\end{equation}
where $\oper{A}$ is either $\iden$ or $\oper{\Pi}$ following the Eq.~\eqref{eq:A-n-revisit}.

Figure~\ref{fig:DisNoise}(c) depicts the Wigner function of $\overline\rho(|\lambda|)$ when $50$ intervention have been applied. The similarity of the Wigner function with the initial state's Wigner function is apparent. Furthermore, thin green lines in Figure~\ref{fig:DisNoise}(d) show $F(\ell)$ for {$200$} noise trajectories that involve intervention, and the thick Green line is the average of them. Using our intervention strategy, obviously resulted in improved fidelity. {One can also observe that perfect fidelity cannot be achieved in our numerical example. This can be seen from the expression Eq.~\eqref{eq:W_r-conv} where one can achieve perfect decoupling only when the effective filter function is a delta distribution, which is not the case in our example. However, it is suggested that our decoupling protocol can modify the states and suppress the noise contribution, reflecting the increase in fidelity.
}

\subsection{Squeezing Noise}\label{sec:numer-squeeze}
In this subsection, we investigate our suppression protocol for {squeezing noise}. We first consider the case where only the {squeezing noise} $\oper{S}(\xi)$ is present, i.e., no displacement noise. We write a similar equation to Eq.~\eqref{eq:dis_state} for the effect of the {squeezing noise} as
\begin{equation}\label{eq:sqz_state}
    \rho(\ell + \mathrm{d}\ell) = \oper{S}(\xi) \rho(\ell) \oper{S}(\xi)^\dagger.
\end{equation}
The controlled state with intervention follows
\begin{equation}
    \rho(\ell + \mathrm{d}\ell) = \oper{S}(\xi) \oper{A} \rho(\ell) \oper{A}^\dagger \oper{S}(\xi)^\dagger
\end{equation}
where here $\oper{A}$ is either $\dfrac{\pi}{2}$ rotation or $-\dfrac{\pi}{2}$ following Eq.~\eqref{eq:A-k-S}.

\begin{figure}[t]
\includegraphics[width=0.95\columnwidth]{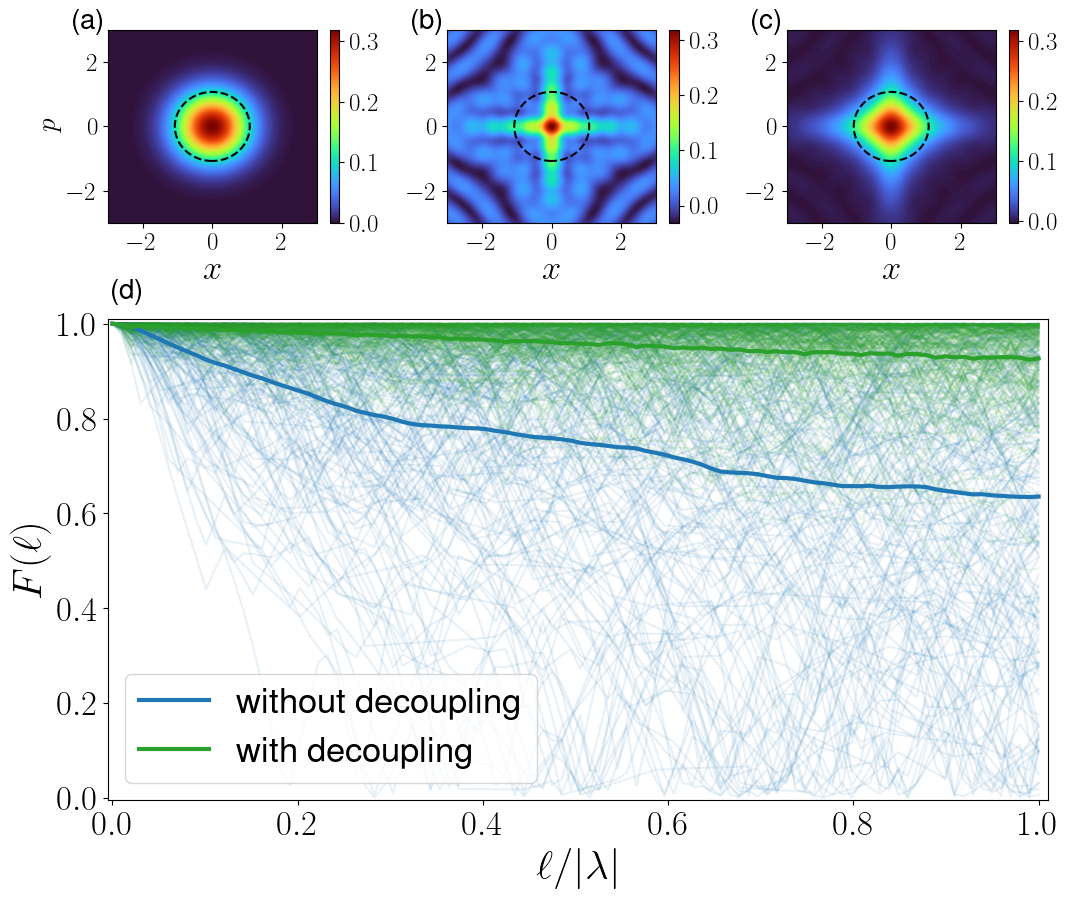}
\caption{\label{fig:SqzNoise} Demonstration of noise suppression of random {squeezing noise}. The details are the same as Figure~\ref{fig:DisNoise}.}
\end{figure}

{For the sake of illustration,} we consider that $\xi$ is a random real number. {This can be easily generalized to complex numbers.} Moreover, similar to displacement noise, we consider CPP for $\xi$ with a rate $0.2$ and a Gaussian pdf with a standard deviation $0.1$. 
Figure~\ref{fig:SqzNoise} shows the result of our numerical calculation for {squeezing noise} and its suppression using our protocol. Similar to the displacement case, top sub-figures---from left to right---show the Wigner function of the initial state, a final state without intervention, and a final state with the noise suppression protocol. Figure~\ref{fig:SqzNoise}(d) shows the fidelity $F(\ell)$ for different noise trajectories without(with) intervention in thin blue(green) lines, and the thick lines are the average fidelity. Similar to displacement noise, our numerical calculation demonstrates that our noise suppression protocol results in improved fidelity.

\begin{figure}[t]
\includegraphics[width=0.95\columnwidth]{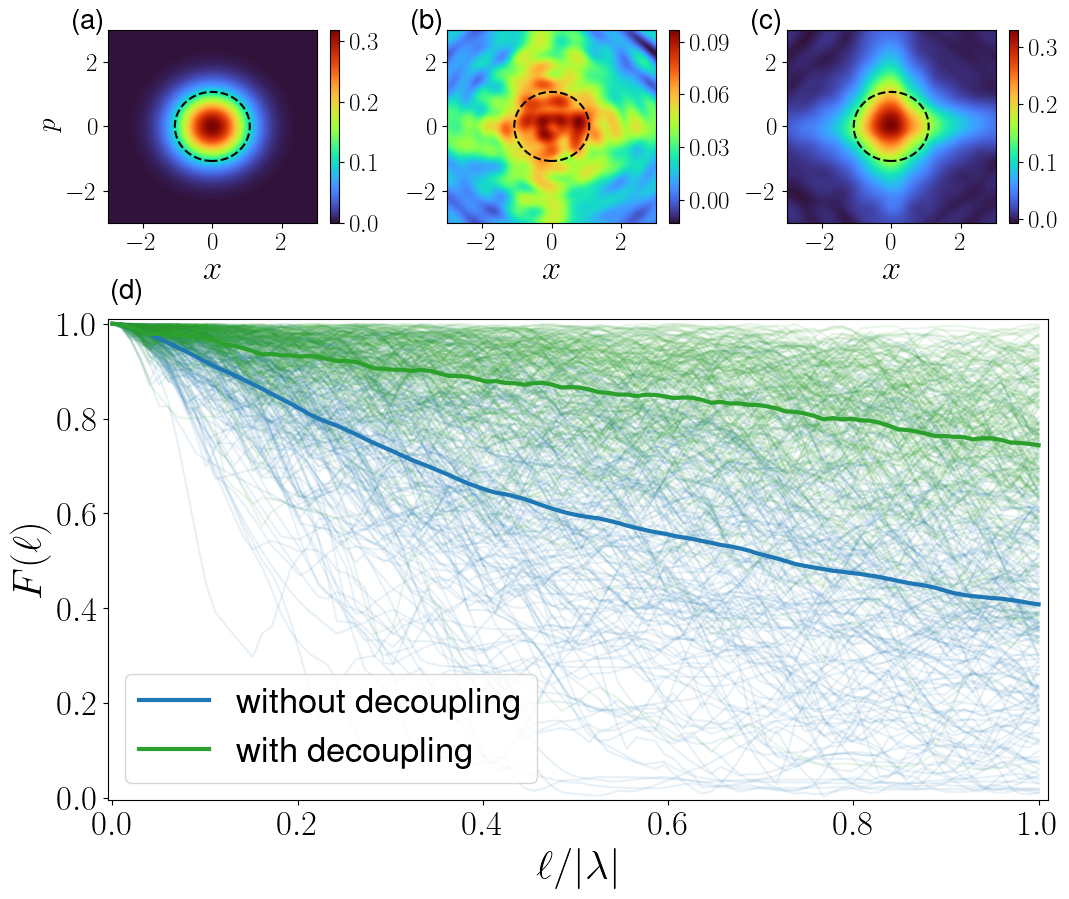}
\caption{\label{fig:combNoise} Demonstration of noise suppression of combined displacement and {squeezing noise}. The details are the same as Figure~\ref{fig:DisNoise}.the}
\end{figure}

Now, we consider the scenario where both displacement and {squeezing noise} are {presented.} {The dynamics} of the system without interventions follows
\begin{equation}\label{eq:DisSq}
    \rho(\ell + \mathrm{d}\ell) = \oper{D}(\zeta)\oper{S}(\xi) \rho(\ell) \oper{S}(\xi)^\dagger \oper{D}(\zeta)^\dagger,
\end{equation}
and {the} dynamics with interventions follows 
\begin{equation}
    \rho(\ell + \mathrm{d}\ell) = \oper{D}(\zeta)\oper{S}(\xi) \oper{A} \rho(\ell) \oper{A}^\dagger \oper{S}(\xi)^\dagger \oper{D}(\zeta)^\dagger.
\end{equation}
In this case, following the discussion in the previous section, we use intervention elements defined in Eq.~\eqref{eq:Gaussian-G}. The result is shown in Fig.~\ref{fig:combNoise}. Because both types of noise are present in this simulation, we lowered the standard deviation by a factor of $\sqrt{2}$; this makes the comparison fair. Again, our numerical shows improvement in Fidelity using our suppression protocol.

\subsection{Effect of number of interventions}
\begin{figure}[t]
\includegraphics[width=0.95\columnwidth]{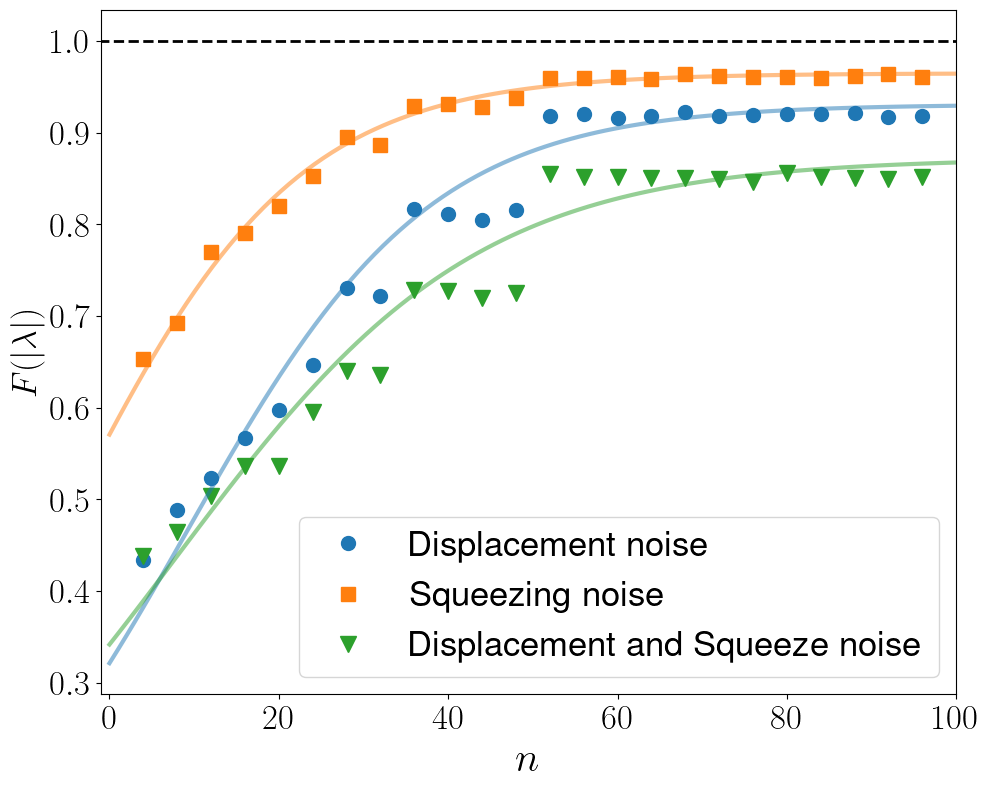}
\caption{\label{fig:FvsN} This shows improvement of fidelity vs. the number of interventions for {Squeezing noise} (orange {squares}), Displacement noise (blue {circles}) and combined displacement and {squeezing noise} (green {triangles}). Data points are the result of the simulation and solid lines are logistic curve fit to data.}
\end{figure}

In all previous numerical demonstrations, we used $n=50$ interventions for each realization of the noise. Here, we run the simulation of our noise suppression protocol for a range of $n$ from $1$ to $100$, using $100$ noise realization, and calculate the average final fidelity, i.e., $F(|\lambda|)$. Our result is shown in Figure~\ref{fig:FvsN} where the points are our numerical simulation, and solid lines are logistic curve fits of the points. As we expect, by increasing the number of interventions, the average fidelity gets closer to its theoretical maximum $1$. Note that we choose a logistic curve fit to guide the eyes and do not provide any analytical reason for such a curve. Finding the actual curve might be useful for experimental consideration of our protocol although we leave this for future investigation.

\section{Conclusions}\label{sec:conclusion}
We considered state transfer problems with continuous variables where the channel by which the state passes is noisy. This problem is inspired by quantum communication by using continuous degrees of freedom in real physical settings, e.g. a photon carrying information via optical fiber or free space communication between satellite and ground stations, where in both cases one can expect a distortion of the signals by the noise from the medium. We modeled such noise as a random mixture of unitary operations. The noisy propagators of the state can be interpreted similarly as random unitary channels, where in this case we employ the path length as a dynamical parameter. We introduced a noise decoupling protocol based on this analogy, where we suppose that the path between the transmitter node and the receiver node can be intervened, allowing us to insert control operations to modify the effective generator of the propagators in such a way that the effect of the noise vanishes, i.e. to achieve an identity channel. 

We provided detailed analyses for the noise channels generated by linear and quadratic polynomials of creation and annihilation operators. We observed that the target state can be recovered for the general noise profile when the random variables do not depend on the path length, while for the case with path-dependent but ergodic and stationary noise, such a situation can be achieved for fast interventions. Furthermore, in principle, a similar analysis can be extended to the case of higher-order polynomial generators. We demonstrated our results numerically, where one can see the increase in fidelity improved by the proposed interventions. This suggests a promising technique of noise decoupling protocols to improve the efficiency of communication tasks in real physical settings, e.g. fiber and free space communications, as previously mentioned.

{
Finally, we remark here that, in our noise model and the similar model literature, the noise comprises two components: the noise generators and the noise profile. The dynamical decoupling technique only deals with the noise generators and analytically avoids the errors caused by the noise characteristics by using the limit procedure and the consideration in an asymptotic regime. In practice, this type of error is important and has been studied in the field of quantum control for several decades in the name of dynamical pulse optimization. A similar limitation of the technique remains in the CV case; we can find a control group for i.e. non-Gaussian noise generators, however, to enable the complete decoupling in the generic scenario, one needs to employ the limit procedure, e.g. the Trotter product-like approximation and has to accept the practical error given by the approximation. In this work, we have shown that the basic principle of dynamical decoupling can be applied and one can find control sets for higher-order noise generators. It is then interesting to consider transferring techniques such as control optimization from dynamical decoupling to noise decoupling in CV systems and develop a practical noise decoupling protocol in CV communication systems. We left these open questions for further studies.
}

\section*{Acknowledgments}
We would like to thank {\L}ukasz Rudnicki, Jan Ko{\l}odynski and Marcin Jarzyna for valuable discussions and suggestions. FS acknowledges support by the Foundation for Polish Science (IRAP project, ICTQT, contract no. 2018/MAB/5, co-financed by EU within Smart Growth Operational Programme). BT acknowledges support of IWY program at DATA61|CSIRO. FS and BT contributed equally to this work.

\bibliographystyle{apsrev4-1}
\bibliography{main.bbl}

\end{document}